\documentclass[12pt,letterpaper]{article}
\usepackage[left=1in,top=1in,right=1in,nohead]{geometry}
\usepackage{enumerate}
\usepackage{graphicx}
\usepackage{amsmath}
\usepackage{amssymb}
\usepackage{bm}
\usepackage{subfig}
\usepackage{customcommands}
\usepackage{jheppub}

\newcommand{\dd}{\mathrm{d}}
\newcommand{\RR}{\mathbb{R}}
\newcommand{\abs}[1]{\left\vert #1 \right\vert}

\usepackage[normalem]{ulem}

\numberwithin{equation}{section}

\title{Quantum chaos and pole skipping in two-dimensional conformal perturbation theory}

\author[1]{Curtis T.~Asplund,}
\author[2]{Sebastian Fischetti,}
\author[3]{Alexandra Miller,}
\author[4,5]{and David M.~Ramirez}

\affiliation[1]{Department of Physics \& Astronomy, San Jos\'e State University, One Washington Square, San Jos\'e, California, 95192}
\affiliation[2]{Department of Physics \& Astronomy, Weber State University, Ogden, Utah, 84401}
\affiliation[3]{Department of Physics \& Astronomy, Sonoma State University, Rohnert Park, California, 94928}
\affiliation[4]{Department of Applied Mathematics and Theoretical Physics, University of Cambridge, Cambridge CB3 0WA, United Kingdom}
\affiliation[5]{CPHT, CNRS, \'Ecole polytechnique, IP Paris, F-91128 Palaiseau, France}

\emailAdd{curtis.asplund@sjsu.edu}
\emailAdd{sfischetti@weber.edu}
\emailAdd{millerale@sonoma.edu}
\emailAdd{david.ramirez@polytechnique.edu}

\abstract{
We analyze pole skipping of stress tensor two-point functions in two-dimensional quantum field theories perturbed away from conformality by a relevant deformation.  The retarded two-point Green's function can be formally computed in conformal perturbation theory, though it results in singular expressions.  We propose a natural interpretation of these expressions and compute the resulting Green's function to leading nontrivial order in the deformation.  As a check of our results, we compare the Lyapunov exponents and butterfly velocities we find from our computed skipped poles to those obtained from both a leading-order conformal field theory analysis using Ward identities, as well as to a holographic gravitational dual perturbed by a massive scalar field; we find precise agreement.  We comment on extensions to sub-leading order, where agreement with holographic expectations would no longer be expected.
}

\begin{document}

\maketitle

\section{Introduction}
\label{sec:intro}

In recent years, chaos has been recognized as an extremely fruitful probe of important properties of many-body systems and quantum field theories.  However, while chaos in classical systems is a well-established field of study, defining chaos in quantum systems is an ongoing area of research, particularly for extended systems such as lattice models and quantum field theories.  One useful diagnostic of quantum chaotic behavior in such extended systems that has emerged in recent years, motivated specifically by holographic gauge/gravity-duality insights, has been out-of-time-ordered correlation functions (OTOCs)~\cite{1969JETP...28.1200L,Almheiri:2013hfa,Shenker:2013pqa, Maldacena:2015waa, Shenker:2013yza, Roberts:2014isa, Shenker:2014cwa, Roberts:2014ifa, Khetrapal:2022dzy, Lin:2018tce, Gu:2016hoy, Das:2022jrr, Plamadeala:2018vsr, Dong:2022ucb}.\footnote{Other attempts at defining chaos in a quantum theory include comparing the spectrum of the system's Hamiltonian with that of random matrices~\cite{Cotler:2016fpe} and semi-classical methods that analyze quantizations of classically chaotic systems (for overviews see, e.g., \cite{berry1987bakerian, cvitanovic2020chaos}).}  These correlation functions can be used to define a quantum Lyapunov exponent~$\lambda_\text{L}$ and butterfly velocity~$v_\text{B}$, characterizing the exponentially-growing effect of the insertion of a local operator:
\be
\ev{W(t,x)V(0)W(t,x)V(0)}_\beta \sim e^{\lambda_\text{L}(t - x/v_\text{B})},
\ee
where~$W$ and~$V$ are simple local operators and the correlator is computed in a thermal state with inverse temperature~$\beta$.

Using OTOCs as a diagnostic of chaos, however, poses a difficulty: they tend to be more challenging to compute than more conventional correlators, both because of the out-of-time-ordering and because of the need to work at nonzero temperature.  In the holographic contexts in which OTOCs have been used extensively, these challenges are broadly circumvented by computing them holographically, which involves studying the backreaction of a shock wave on the bulk geometry~\cite{Maldacena:2015waa,Shenker:2013yza}.  But for non-holographic theories, the need to compute OTOCs directly is an apparent challenge.

Fortunately, there is evidence that, at least to some extent, the diagnostic information obtained from OTOCs -- namely, the Lyapunov exponent and butterfly velocity -- can be inferred from the so-called ``pole-skipping'' points of the retarded stress tensor two-point function~\cite{Grozdanov:2017ajz, Blake:2017ris, Blake:2018leo, Blake:2019otz, Grozdanov:2018kkt, Natsuume:2019sfp, Natsuume:2019xcy, Natsuume:2019vcv, Ahn:2019rnq, Wu:2019esr, Ramirez:2020qer, Blake:2021hjj, Yadav:2023hyg, Baishya:2023mgz, Wang:2022mcq, Ning:2023ggs, Grozdanov:2023txs, Ahn:2024gjh, Natsuume:2023lzy, Natsuume:2023hsz, Knysh:2024asf, Saha:2024bpt,Baishya:2024gih, Liu:2020yaf, Ahn:2025exp, Jeong:2021zhz, Yuan:2024utc, Lilani:2025wnd}.  For a simple summary, consider the case of a two-dimensional theory with retarded two-point function of the energy density~$G_{T^{tt} T^{tt}}(\omega,k)$.  When analytically continuing~$(\omega,k)$ to~$\CC^2$, one can define the surface on which~$G_{T^{tt} T^{tt}}(\omega,k)$ vanishes, as well as a surface on which it diverges.  The intersection of these two surfaces in~$\CC^2$ identifies special values~$(\omega_*,k_*)$ at which~$G_{T^{tt} T^{tt}}(\omega_*,k_*)$ is ill-defined; these special values are referred to as the skipped poles.  It has been observed that at least in holographic theories, the  skipped pole with $\mathrm{Im}(\omega)>0$ seems to precisely correspond to the Lyapunov exponent and butterfly velocity defined through the OTOC via the identifications
\be
\lambda_\text{L} = -i\omega_*, \quad v_\text{B} = \frac{\omega_*}{k_*}.
\label{eq:lambda-vb-def}
\ee
It is now understood that this identification holds in any theory with a holographic dual~\cite{Chua:2025vig}.

Whether or not this matching between skipped poles and chaos is universal, however, depends on whether the connection between skipped poles and chaos exponents holds away from holographic theories.  For example, as suggested in~\cite{Choi:2020tdj}, one might expect from the holographic results that~$-i\omega_*$ and~$\omega_*/k_*$ merely give the stress-tensor contribution to the Lyapunov exponent and butterfly velocity\footnote{For this reason, we will therefore refer to~$\omega_*/k_*$ as a ``pole-skipping velocity''~$v_\mathrm{PS}$ to differentiate it from the genuine butterfly velocity~$v_\text{B}$ defined by the OTOC.}.  It is therefore desirable to explore the connection between skipped poles and Lyapunov exponents away from holography (and away from conformality).

The purpose of this paper is to take a first step in this direction by computing the skipped poles of a two-dimensional conformal field theory (CFT) perturbed by a relevant scalar deformation with holomorphic conformal weight~$h \in (0,1)$.  While the leading-order perturbation to the retarded Green's function and skipped poles can be obtained using Ward identities, as we demonstrate in Section~\ref{sec:Ward}, proceeding to higher orders using conformal perturbation theory introduces formal expressions containing ill-defined, singular integrals.  One of the main contributions of this paper is to explain how to interpret the ill-defined expressions distributionally, demonstrating that an explicit calculation of the integrals appearing at leading order reproduces the leading singular behavior of the results obtained from Ward identities.  Specifically, we use conformal perturbation theory to obtain explicit results for the Euclidean Green's function for all~$h \in (0,1)$, and for the perturbation to the skipped poles when~$h = 1/2$ and~$h = 1$. 

As we discuss further in the main text, the correlation functions of interest, even at the unperturbed CFT, are distributional objects and require a prescription for dealing with ambiguities at coincident points. The resulting contact terms, which manifest in momentum space as polynomials in $\omega$ and $k$, need not agree between different natural regularization procedures, and indeed the correlation functions we find in Section~\ref{sec:Ward}, obtained by studying the Ward identities of the generating functional on curved backgrounds, will differ from direct Fourier transform of standard CFT$_2$ results (an analogous distinction was pointed out in \cite{Policastro:2002tn})\footnote{A recent discussion of distributional aspects of correlation functions can be found in \cite{Arnaudo:2026der}. We would like to thank the authors of this work for discussions on this topic.}. Ultimately, these differences will not play a role in our investigation of pole skipping, but it would be nice to sort out these technical issues in future work and ensure they remain harmless at higher orders in perturbation theory.

It turns out that the leading-order correction to the Green's function is of order~$1/c$, where~$c$ is the central charge of the background CFT.  Therefore, although we make no assumption about our theory having a holographic dual, we expect that the subleading correction to the Green's function should match that of a holographic theory.  To corroborate this expectation, we additionally compute the Lyapunov exponent and butterfly velocity of a holographic theory deformed away from conformality, and we find exact agreement with the results obtained using Ward identities and conformal perturbation theory. We also anticipate that a direct bulk calculation of the full retarded Green's function will agree with the CFT analysis, and indeed the near-horizon analysis exhibits the matching in \eqref{eq:lambda-vb-def} \cite{Chua:2025vig}; however, the bulk modes governing $G_{T_{\mu\nu}T_{\rho\sigma}}$ (i.e., solutions to the linearized equations of motion) are largely gauge artifacts due to the lack of propagating gravitons in $2+1$ dimensions, complicating the standard bulk analysis. We leave a more complete decoupling and solution of the bulk system for future work and consider pole skipping purely from the boundary perspective.

This paper is organized as follows.  In Section~\ref{sec:Ward}, we describe how to obtain the leading-order correction to the skipped poles (and pole-skipping velocity) of the deformed CFT using Ward identities.  In Section~\ref{sec:CFT}, we then review how to use conformal perturbation theory to obtain formal expressions for the leading-order correction to the Euclidean stress tensor Green's function in the deformed theory.  In Section~\ref{sec:integrals}, we discuss how to properly interpret the formal expressions in terms of homogeneous distributions, allowing us to evaluate them explicitly.  We then use the result to compute the retarded Lorentzian Green's function, allowing us to extract the perturbation to the skipped poles.  In Section~\ref{sec:gravity}, we perform a holographic calculation of the chaos exponents in a deformed holographic CFT.  We find that the results obtained using Ward identities, conformal perturbation theory, and holographic approaches all agree.  We conclude in Section~\ref{sec:conclusion} by outlining next steps, specifically a CFT computation of the perturbed OTOCs or sub-subleading corrections to the skipped poles.

\section{Leading-Order Correction from Ward Identities}
\label{sec:Ward}

In this section, we determine the leading-order correction to pole skipping by exploiting the diffeomorphism and dilatation Ward identities, as well as thermodynamics of the perturbed CFT.  We consider a CFT deformed by 
\begin{equation}
  S = S_0 + \int \dd^2 x \, \lambda(x) {\cal O}(x), 
\end{equation}
where ${\cal O}(x)$ is a relevant scalar operator of the bare CFT, which for
simplicity we take to be a scalar Virasoro primary with conformal dimension $\Delta=2h$. Here we have allowed for a spatially varying coupling $\lambda(x)$, as it proves useful in deriving the Ward identities we turn to next.

As shown in \cite{Davison:2024msq}, the combination of the constrained kinematics in~$1+1$ dimensions and diffeomorphism invariance fixes every stress tensor retarded two-point function~$G_{\mu\nu,\rho\sigma} \equiv G_{T_{\mu\nu}T_{\rho\sigma}}$ in terms of a single unknown function, which we can take to be the correlation function~$G_{\Theta\Theta}$ of the trace~$\Theta \equiv T^\mu{}_\mu$.\footnote{A note on notation: We use $G_{\mu\nu,\rho\sigma}$ and $G_{\Theta\Theta}$ for what is called ${-}G_R^{T_{\mu\nu}T_{\rho\sigma}}$ and ${-}G_R^\text{trace}$ in \cite{Davison:2024msq}, where the negative sign is due to a different convention for the definition of retarded Green's functions.}  
To summarize the argument, diffeomorphism invariance leads to the Ward  identity\footnote{Throughout this work, we assume that there are no gravitational anomalies.} $\nabla_\nu \langle T^{\mu\nu}(x)\rangle = {-}\langle {\cal O}(x) \rangle \nabla^\mu \lambda$, which can be further differentiated with respect to the metric to yield a constraint on the two-point functions $G_{\mu\nu,\rho\sigma}$.  Evaluating the result on a flat background and Fourier transforming yields
\begin{align}
  p^\mu G_{\mu\nu,\rho\sigma}(\omega,k) ={}& p^\mu \left(\langle T_{\mu\rho} \rangle \eta_{\nu\sigma} + \langle T_{\mu\sigma}\rangle \eta_{\nu\rho} - \langle T_{\rho\sigma}\rangle \eta_{\mu\nu}\right) .
\end{align}
Using this identity, one finds that any $G_{\mu\nu,\rho\sigma}$ can be expressed in terms of $G_{\Theta \Theta}$.  For example, introducing light-cone coordinates $x^\pm = x \pm t$, the two-point function of~$T_{--}$ is
\begin{align}
\label{eq:GRpppp}
  G_{--,--}(\omega,k) = \frac{1}{16}\left(\frac{\omega+k}{\omega - k}\right)^2 G_{\Theta\Theta} + \frac{1}{4}\frac{\omega + k}{\omega - k} (\varepsilon+p),
\end{align}
where~$\varepsilon = \langle T^{tt}\rangle$ and~$p = \langle T^{xx} \rangle$ are the energy density and pressure, respectively.  Although discussions of pole skipping often focus on the energy density two-point function~$G_{tt,tt}$, comparison with Euclidean calculations in the rest of this paper makes it more natural to work with~$G_{--,--}$, since the Euclidean two-point function of the holomorphic stress tensor that we compute in Sections~\ref{sec:CFT} and~\ref{sec:integrals} then continues to the real time two-point function $G_{--,--}$.

Next, the dilatation Ward identity relates $G_{\Theta\Theta}$ to correlation functions of the deforming operator ${\cal O}$.  Weyl invariance yields the Ward identity
\begin{align}
  \langle \Theta(x) \rangle ={}& {-}2(1-h) \lambda \langle {\cal O}(x)\rangle + \frac{c}{24\pi} R,
\end{align}
where $R$ is the scalar curvature of the background metric.  This again leads to constraints on the two-point functions by differentiation, which ultimately allows us to relate $G_{\Theta\Theta}$ to $G_{{\cal O}{\cal O}}$ and the one-point function $\langle {\cal O}\rangle$:
\begin{align}
\label{eq:GROO}
  G_{\Theta\Theta}(\omega,k) ={}& \frac{c}{12\pi} (\omega^2 - k^2) + 4(1-h)^2 \left(\lambda^2 G_{{\cal O}{\cal O}}(\omega,k) - \frac{h}{1-h} \lambda \langle{\cal O}\rangle \right)\, .
\end{align}

Combining~\eqref{eq:GRpppp} and~\eqref{eq:GROO} allows us to compute~$G_{--,--}$ directly from the two-point and one-point functions~$G_{\Ocal\Ocal}$ and~$\langle \Ocal \rangle$, yielding:
\begin{align}
    G_{--,--}(\omega,k) ={}& \frac{c}{192\pi} \frac{(\omega+k)^3}{\omega - k} + \frac{1}{4} \frac{\omega+k}{\omega-k} (\varepsilon+p) \nonumber \\
    & + \frac{(1-h)^2}{4} \left(\frac{\omega+k}{\omega-k}\right)^2 \left(\lambda^2 G_{\Ocal \Ocal}(\omega, k) - \frac{h}{1-h} \, \lambda \langle \Ocal \rangle \right)\, .
    \label{eq:GR-full}
\end{align}
We emphasize that this relationship is non-perturbative: it applies for general~$\lambda$.  However, since~$G_{\Ocal\Ocal}$ and~$\langle \Ocal \rangle$ each depend non-trivially on~$\lambda$, to obtain~$G_{--,--}$ to a given order in~$\lambda$ we will compute~$G_{\Ocal\Ocal}$ and~$\langle \Ocal \rangle$ in conformal perturbation theory.  Here we restrict our analysis to leading nontrivial order: the~$O(\lambda^2)$ behavior of~$G_{--,--}$ can be obtained from the unperturbed two-point function~$G_{0,\Ocal\Ocal}$, the~$O(\lambda)$ correction to~$\langle \Ocal \rangle$, and the~$O(\lambda^2)$ correction to the enthalpy~$\varepsilon+p$.  Writing
\be
G_{--,--} = G_{0,--,--} + \lambda^2 G_{2,--,--} + O(\lambda^3),
\ee
we find that in the unperturbed CFT at temperature~$\beta^{-1}$,
\be
G_{0,--,--} = \frac{c}{192\pi} \frac{\omega + k}{\omega - k} \left[\left(\omega+k\right)^2 + \left(\frac{4 \pi}{\beta}\right)^2 \right] \, , \label{eq:GR0-WI}
\ee
using the familiar CFT thermodynamic relation $\varepsilon+p= \pi c/(3\beta^2)$.  Proceeding to subleading order, we use the corrections to~$\varepsilon+p$ and~$\langle \Ocal \rangle$ as well as the unperturbed CFT result~$G_{0,\Ocal\Ocal}$ provided in Appendix~\ref{app:cpt-thermo} to obtain an expression for~$G_{2,--,--}$ given explicitly in~\eqref{eq:GR2-WI}. 

With this leading correction to $G_{--,--}$, we now turn to the effect on the location of the skipped poles. To do so, we first write
\begin{align}
    \omega_* ={}& \omega_0 + \lambda^2 \omega_2 + O(\lambda^3)\, , & k_* ={}& k_0 + \lambda^2 k_2 + O(\lambda^3)\, ,
\end{align}
where $\omega_0 = k_0 = 2\pi i/\beta$ is the unperturbed pole skipping location\footnote{The attentive reader may recognize that the unperturbed result \eqref{eq:GR0-WI} only exhibits pole skipping at $\omega=k$, while the energy density has skipped poles at $\omega= \pm k$. The missing skipped pole is found in the correlator $G_{++,++}$, which can be analyzed in exactly the same manner as $G_{--,--}$.}, as can be seen from \eqref{eq:GR0-WI}. We then solve~$G_{--,--}(\omega_*,k_*) = 0$ order-by-order in~$\lambda$ for the coefficients~$\omega_i$ as a function of the~$k_i$ in order to identify the location of the perturbed zero surface of~$G_{--,--}$, and we solve~$1/G_{--,--}(\omega_*,k_*) = 0$ order-by-order for the~$\omega_i$ as a function of the~$k_i$ to identify the singular surface of~$G_{--,--}$\footnote{This procedure for finding the shift in the singular surface is effectively performing a resummation of the leading-order perturbative result for $G_{--,--}$.}.  The perturbed skipped poles are then found by identifying where these surfaces intersect, and the corresponding pole-skipping velocity is $v_\text{PS}= \omega_*/k_*$.  This procedure ultimately leads to
\begin{align}
  v_\text{PS} 
  ={}& 1 + \frac{3 \pi^{4h}}{8} \frac{8 (1 -2h) \csc(2\pi h) - 4\pi
       h (1 - h) \csc^2(\pi h)}{
       \left[\Gamma \left(1 - h\right) \Gamma \left(\frac{1}{2}+h\right)\right]^2 } \frac{\lambda^2}{c \beta^{4(h-1)}} + \dotsb\, . \label{eq:vPSWard}
\end{align}

While this approach has quickly lead us to the leading correction to the pole skipping velocity (which we will see in Sec.~\ref{sec:gravity} is reproduced holographically), carrying out the perturbative expansion to higher orders requires directly evaluating the integrals arising in conformal perturbation theory. These integrals require regularization and some care to evaluate, so in the next two sections we will take this direct approach to reproduce the results in this section, leaving the higher order corrections to future work. For an extended discussion of pole skipping along the lines of this section, we encourage the reader to consult \cite{Davison:2025xdj}.

\section{Conformal Perturbation Theory}
\label{sec:CFT}

In this section, we use conformal perturbation theory to determine the leading order correction to the stress tensor Green's function in the deformed theory. 
This correction is determined by correlation functions in the undeformed (bare) CFT. 
For a review of conformal perturbation theory see, e.g., \cite{Amoretti:2017aze}, or \cite{Berenstein:2014cia, Berenstein:2016avf} for discussions of it in an AdS/CFT context.
We will work in Euclidean signature using the following conventions.  The thermal Euclidean cylinder,~$S_\beta^1 \times \RR$, will often be denoted by a shorthand ``cyl'', and complex coordinates on the cylinder will be denoted by a tilde:~$\tilde{z}$.  We will often conformally map the cylinder to the plane using the map~$\tilde{z} = (\beta/2\pi) \ln z$, where untilded coordinates like~$z$ are taken to live on the complex plane\footnote{Note that this is not the conformal transformation used in the standard CFT$_2$ radial quantization: the constant radius circles in the plane are mapped to the thermal circle of radius $\beta$.}.  When integrating on the thermal cylinder or complex plane, we will use the shorthand
\be
d^2 z \equiv \frac{d\bar{z} \wedge dz}{2i} = dx \wedge dy,
\ee
where as usual~$z = x + iy$.  When possible, we will use tildes (and lack thereof) to indicate whether an integral is taken over the thermal cylinder or over the plane.

\subsection{Perturbative expansion of Euclidean Green's function}
\label{sec:PertExp}

If $S_0$ is the action of our bare CFT, the deformed theory we consider has the Euclidean action
\begin{equation}
  \label{eq:CPT-action}
  S = S_0 + \lambda \int_{\mathrm{cyl}} \dd^2 \tilde{z}\, \sqrt{g(\tilde{z}, \bar{\tilde{z}})} \, {\cal O}(\tilde{z}, \bar{\tilde{z}})\, , 
\end{equation}
where as mentioned above ${\cal O}(x)$ is a relevant scalar operator of the bare CFT with conformal dimension $\Delta = 2h$ (i.e.~with holomorphic weight
$h$), with~$h \in (0,1)$.  Note that we will frequently suppress the dependence on the antiholomorphic coordinate $\bar{\tilde{z}}$.  Explicitly, we will assume that the partition function of the deformed theory is given by
\begin{equation}
  \label{eq:Zlambda-def}
  Z[\lambda,g] = Z^{(0)}[g] \left\langle \exp \left[ {-} \lambda \int_{\mathrm{cyl}} \dd^2 \tilde{z}\, \sqrt{g(\tilde{z})} {\cal O}(\tilde{z}) \right] \right\rangle_0 \equiv Z^{(0)}[g] Z^{(\lambda)}[\lambda,g]\, .
\end{equation}
Here $Z^{(0)}[g]$ is the partition function of the unperturbed CFT on a background
with metric $g$ and the angular brackets $\langle \cdot \rangle_0$
indicate a (normalized) correlation function in the bare CFT.

Slightly more convenient for our purposes will be the generating
functional of connected correlators, $W[\lambda,g]$, defined as
\begin{equation}
  \label{eq:Wlambda-def}
  W[\lambda, g] = {-} \log Z[\lambda,g] = W^{(0)}[g] + W^{(\lambda)}[\lambda,g]\, .
\end{equation}
We have also introduced $W^{(0)} = {-} \log Z^{(0)}$ and
$W^{(\lambda)} = {-} \log Z^{(\lambda)}$. Expanding the exponential in
\eqref{eq:Zlambda-def} lets us write $Z^{(\lambda)}$ and $W^{(\lambda)}$
in terms of correlation functions in the bare CFT. To quadratic order
in $\lambda$, one finds:
\begin{subequations}
\begin{align}
  Z^{(\lambda)} ={}& 1 - \lambda \int \dd^2 \tilde{z}\, \sqrt{g(\tilde z)} \langle {\cal O}(\tilde{z}) \rangle_0 + {\lambda^2 \over 2} \int \dd^2 \tilde{z}_1 \, \dd^2 \tilde{z}_2\, \sqrt{g(\tilde{z}_1) g(\tilde{z}_2)} \langle {\cal O}(\tilde{z}_1) {\cal O}(\tilde{z}_2) \rangle_0 + \dotsb \label{eq:Zlambda-exp}, \\
  W^{(\lambda)} ={}& \lambda \int \dd^2 \tilde{z}\, \sqrt{g(\tilde z)} \langle {\cal O}(\tilde{z}) \rangle_0 - {\lambda^2 \over 2} \int\dd^2 \tilde{z}_1 \, \dd^2 \tilde{z}_2\, \sqrt{g(\tilde{z}_1) g(\tilde{z}_2)} \langle {\cal O}(\tilde{z}_1) {\cal O}(\tilde{z}_2) \rangle_0^{(c)} + \dotsb , \label{eq:Wlambda-exp}
\end{align}
\end{subequations}
where the superscript $(c)$ indicates a connected correlation function,
e.g.~$\langle{\cal O}_1 {\cal O}_2\rangle^{(c)} = \langle {\cal O}_1
{\cal O}_2\rangle - \langle{\cal O}_1\rangle \langle{\cal
  O}_2\rangle$.

We now use these definitions to write down formal expressions for the stress tensor correlation functions in the deformed theory. The essential fact we use is that inserting a stress tensor into a
correlation function can be obtained by differentiating with respect
to the metric:
\begin{equation}
  {\delta \over \delta g^{\mu\nu}(\tilde{u})} \left[\sqrt{g_1 \dotsb g_n} \langle {\cal O}_1 \dotsb {\cal O}_n\rangle^{(c)} \right] = {-} {1 \over 2} \sqrt{g(\tilde{u})g_1 \dotsb g_n} \langle T_{\mu\nu}(\tilde{u}) {\cal O}_1 \dotsb {\cal O}_n\rangle^{(c)}\, . \label{eq:dg-corr}
\end{equation}
In particular, the one- and two-point functions can be found by
differentiating $W$:
\begin{align}
  \sqrt{g(\tilde{u})} \langle T_{\mu\nu}(\tilde{u}) \rangle ={}& 2 {\delta W \over \delta g^{\mu\nu}(\tilde{u})}\, , & \sqrt{g(\tilde{u}) g(\tilde{v})} \langle T_{\mu\nu}(\tilde{u}) T_{\rho\sigma}(\tilde{v}) \rangle^{(c)} ={}& {-} 4 {\delta^2 W \over \delta g^{\mu\nu}(\tilde{u}) \delta g^{\rho\sigma}(\tilde{v})}\, . \label{eq:TEV-def}
\end{align}
Combining equations \eqref{eq:Wlambda-exp}, \eqref{eq:dg-corr}, and
\eqref{eq:TEV-def} leads to the one-point function
\begin{align}
  \langle T_{\mu\nu}(\tilde{u})\rangle ={}& \langle T_{\mu\nu}(\tilde{u})\rangle_0 - \lambda \int\dd^2 \tilde{z} \sqrt{g(\tilde{z})} \langle T_{\mu\nu}(\tilde{u}) {\cal O}(\tilde{z})\rangle_0^{(c)} \nonumber \\
                                  &+ {\lambda^2 \over 2} \int\dd^2 \tilde{z}_1 \dd^2 \tilde{z}_2 \sqrt{g(\tilde{z}_1) g(\tilde{z}_2)} \langle T_{\mu\nu}(\tilde{u}) {\cal O}(\tilde{z}_1) {\cal O}(\tilde{z}_2) \rangle_0^{(c)} + \dotsb\, , \label{eq:TEV-exp}
\end{align}
which can be further differentiated (via \eqref{eq:dg-corr}) to find the
connected two-point function (Euclidean Green's function) $G^\text{E}_{\mu\nu,\rho\sigma}(\tilde{u}, \tilde{v}) \equiv \langle
T_{\mu\nu}(\tilde{u}) T_{\rho\sigma}(\tilde{v})\rangle^{(c)}$:
\be
G^\text{E}_{\mu\nu,\rho\sigma}(\tilde{u}, \tilde{v}) = G^{\text{E}}_{0,\mu\nu,\rho\sigma}(\tilde{u}, \tilde{v}) + \lambda G^{\text{E}}_{1,\mu\nu,\rho\sigma}(\tilde{u}, \tilde{v}) + \lambda^2 G^{\text{E}}_{2,\mu\nu,\rho\sigma}(\tilde{u}, \tilde{v}) + \dotsb \label{eq:GE-exp},
\ee
where
\begin{subequations}
\begin{align}
  G^{\text{E}}_{0,\mu\nu,\rho\sigma}(\tilde{u}, \tilde{v}) ={}& \langle T_{\mu\nu}(\tilde{u}) T_{\rho\sigma}(\tilde{v}) \rangle_0^{(c)}\, , \label{eq:GE0-def}\\
  G^{\text{E}}_{1,\mu\nu,\rho\sigma}(\tilde{u},\tilde{v}) ={}& {-} \int \dd^2 \tilde{z} \sqrt{g(\tilde{z})} \langle T_{\mu\nu}(\tilde{u}) T_{\rho\sigma}(\tilde{v}) {\cal O}(\tilde{z})\rangle_0^{(c)}\, , \label{eq:GE1-def}\\
  G^{\text{E}}_{2,\mu\nu,\rho\sigma}(\tilde{u},\tilde{v}) ={}& {1 \over 2} \int\dd^2 \tilde{z}_1 \dd^2 \tilde{z}_2 \sqrt{g(\tilde{z}_1) g(\tilde{z}_2)} \langle T_{\mu\nu}(\tilde{u}) T_{\rho\sigma}(\tilde{v}){\cal O}(\tilde{z}_1) {\cal O}(\tilde{z}_2) \rangle_0^{(c)}\, . \label{eq:GE2-def}
\end{align}
\end{subequations}
Formal expressions for the higher order corrections can be readily
obtained, but this will suffice for our purposes.  For explicit
evaluation, it will be prove convenient to write these corrections in
terms of $\widehat{{\cal O}} = {\cal O} - \langle {\cal O}\rangle_0$ and
$\widehat{T}_{\mu\nu} = T_{\mu\nu} - \langle T_{\mu\nu}\rangle_0$.  The
connected correlation functions can then be written as
\begin{subequations}
\begin{align}
  \langle T_{\mu\nu}(\tilde{u}) T_{\rho\sigma}(\tilde{v}) {\cal O}(\tilde{z})\rangle_0^{(c)} ={}& \langle \widehat{T}_{\mu\nu}(\tilde{u}) \widehat{T}_{\rho\sigma}(\tilde{u}) \widehat{{\cal O}}(\tilde{z})\rangle_0\, , \\
  \langle T_{\mu\nu}(\tilde{u}) T_{\rho\sigma}(\tilde{v}){\cal O}(\tilde{z}_1) {\cal O}(\tilde{z}_2) \rangle_0^{(c)} ={}& \langle \widehat{T}_{\mu\nu}(\tilde{u}) \widehat{T}_{\rho\sigma}(\tilde{v}) \widehat{{\cal O}}(\tilde{z}) \widehat{{\cal O}}(\tilde{z}_2)\rangle_0 \\
  &- \langle \widehat{T}_{\mu\nu}(\tilde{u}) \widehat{T}_{\rho\sigma}(\tilde{v}) \rangle \langle \widehat{{\cal O}}(\tilde{z}_1) \widehat{{\cal O}}(\tilde{z}_2)\rangle_0 \nonumber \\
                &- \langle \widehat{T}_{\mu\nu}(\tilde{u}) \widehat{{\cal O}}(\tilde{z}_1)\rangle \langle \widehat{T}_{\rho\sigma}(\tilde{v})  \widehat{{\cal O}}(\tilde{z}_2)\rangle_0 \nonumber \\
                &- \langle \widehat{T}_{\mu\nu}(\tilde{u}) \widehat{{\cal O}}(\tilde{z}_2)\rangle \langle \widehat{T}_{\rho\sigma}(\tilde{v}) \widehat{{\cal O}}(\tilde{z}_1) \rangle_0\, . \nonumber
\end{align}
\end{subequations}
Now that we have reduced the computation to evaluating correlation
functions in the unperturbed CFT, we will drop the subscript $0$ on
the expectation values.

We note that we will work exclusively throughout this article via the na\"ive perturbative framework outlined above. In particular, we will not explicitly consider any renormalization group improvements, e.g., solving the appropriate Callan-Symanzik equation for $G^\text{E}_{\mu\nu, \rho\sigma}$ (and we note that the stress-tensor itself is protected from an anomalous dimension via the translation invariance Ward identities \cite{Zamolodchikov:1987ti,Cappelli:1989yu}), as a simple resummation of the leading order result will suffice for our purposes. Furthermore, the pole skipping behavior we are seeking, arising at timescales of order $\beta$, is not a late-time feature and so we will find that we avoid the known breakdown of perturbation theory at late times \cite{Davison:2024msq}.

Finally, because (as discussed in Section~\ref{sec:Ward}) all stress tensor correlators are controlled by the single function~$G_{\Ocal\Ocal}$, we expect all correlators of stress tensor components to exhibit the same pole structure.  We thus restrict our focus largely to the behavior of the holomorphic component of the stress tensor $T \equiv 2\pi T_{zz}$, which continues to the retarded two-point correlator~$G_{--,--}$. 

\subsection{Transforming from the cylinder to the plane and
  evaluating the necessary correlation functions}

In order to evaluate the integrals required for $G^{\text{E}}_1$ and
$G^{\text{E}}_2$, we use conformal invariance to relate correlation
functions on the cylinder $S^1_\beta \times \RR$ to those on the
plane. Once on the plane, the correlation functions can be evaluated
using the conformal Ward identities and the canonical two-point
function for scalar conformal primaries on the plane\footnote{Throughout this subsection we use various standard results in two-dimensional CFTs, which can be found in, e.g., \cite{DiFrancesco:1997nk}.}:
\begin{align}
  \langle {\cal O}(z_1,\bar{z}_1) {\cal O}(z_2,\bar{z}_2) \rangle ={}& {1 \over \abs{z_1 - z_2}^{2\Delta}}\, . \label{eq:OOEV}
\end{align}
Using the conformal map $\tilde{z}(z) = {\beta \over 2\pi} \log z$, the transformation of Virasoro primary operators ${\cal O}_i$ takes the simple form
\begin{equation}
  \langle {\cal O}_1 (\tilde{z}_1, \bar{\tilde{z}}_1) \dotsb {\cal O}_n (\tilde{z}_n, \bar{\tilde{z}}_n)\rangle = \left[\prod_i (\partial_{z_i}\tilde{z}_i)^{{-}h_i} (\partial_{\bar{z}_i}\bar{\tilde{z}}_i)^{{-}\bar{h}_i}\right]\langle {\cal O}_1 (z_1, \bar{z}_1) \dotsb {\cal O}_n (z_n, \bar{z}_n)\rangle\, .
\end{equation}
However, the holomorphic
component of the stress tensor $T$ is not a Virasoro primary (though it is a global
conformal primary), so we will also need the transformation law
\begin{equation}
  T_{\cyl}(\tilde{z}) = (\partial_z \tilde{z})^{{-}2} \left[ T_{\CC}(z) - {c \over 12} \{\tilde{z}(z), z\} \right]\, , 
\end{equation}
where $\{w(z),z\}$ is the Schwarzian derivative:
\begin{equation}
  \label{eq:schwarzian}
  \{w(z), z\} = \partial_z \left({\partial_z^2 w \over \partial_z w^2} \right) - {1 \over 2} \left({\partial_z^2 w \over \partial_z w^2} \right)^2\, .
\end{equation}
Similar results of course hold for the antiholomorphic components
$\bar T \equiv T_{\bar z \bar z}$. 

For the conformal transformation at hand, these rules amount to the
replacements
\begin{align}
  {\cal O}_i(\tilde{z}_i) \leadsto{}& \left({2\pi \over \beta} \right)^{\Delta_i} {\cal O}(z_i)\, , & T_{\cyl}(\tilde{z}) \leadsto{}& \left({2\pi \over \beta} \right)^{2} \left[z^2 T_\CC(z) - {c \over 24} \right] \, .
\end{align}
From the latter result, we note that $\widehat{T}$ is in fact covariant
under the conformal transformation $\tilde{z}\to z$, as
$\langle T_\CC \rangle = 0$:
\begin{align}
  \widehat{T}_{\cyl}(\tilde z) ={}& T_{\cyl}(\tilde{z}) - \langle T_{\cyl}\rangle_0 = \left({2\pi \over \beta} \right)^{2} z^2 T_\CC(z)\, .
\end{align}
Furthermore, on the plane the one-point function of any Virasoro
primary vanishes, i.e.~${\cal O} = \widehat{{\cal O}}$. The conformal Ward
identities then further imply $\langle T \dotsb T {\cal O}\rangle = 0$ for
any number of $T$ insertions. In the following, we will suppress the
labels $\cyl$ and $\CC$, again relying on the coordinates to indicate
whether we are on the cylinder or on the plane.

Together, these results dramatically simplify the corrections to
$G^\text{E} \equiv G^\text{E}_{zz,zz}$. First, we note that the integrand in
\eqref{eq:GE1-def} vanishes, as
\begin{align}
  \langle T(\tilde{u}) T(\tilde{v}) {\cal O}(\tilde{z})\rangle^{(c)} ={}& \langle \widehat{T}(\tilde{u}) \widehat{T}(\tilde{v}) \widehat{{\cal O}}(\tilde{z})\rangle_0 \propto \langle T(u) T(v) {\cal O}(z)\rangle = 0\, .
\end{align}
Hence~$G^\text{E}_1 = 0$.  Thus the first non-trivial correction to $G^\text{E}$ arises at order
$\lambda^2$ and is given by
\begin{align}
  G^{\text{E}}_2(\tilde{u},\tilde{v}) ={}& {1 \over 2} \int\dd^2 \tilde{z}_1 \, \dd^2 \tilde{z}_2 \left[\langle \widehat{T}(\tilde{u}) \widehat{T}(\tilde{v}) \widehat{{\cal O}}(\tilde{z}_1) \widehat{{\cal O}}(\tilde{z}_2) \rangle_0 - \langle \widehat{T}(\tilde{u}) \widehat{T}(\tilde{v}) \rangle \langle \widehat{{\cal O}}(\tilde{z}_1) \widehat{{\cal O}}(\tilde{z}_2) \rangle_0\right] \nonumber \\
  ={}& {1 \over 2} \left({2\pi \over \beta} \right)^{4h} u^2 v^2 I(u,v)\, , \label{eq:GE2int}
\end{align}
where we have introduced the integral
\begin{align}
  I(u,v) \equiv \int {\dd^2 z_1 \, \dd^2 z_2  \over \abs{z_1 z_2}^{2(1-h)}}\left[\langle T(u) T(v) {\cal O}(z_1) {\cal O}(z_2) \rangle_0 - \langle T(u) T(v) \rangle_0 \langle {\cal O}(z_1) {\cal O}(z_2) \rangle_0\right]\, . \label{eq:Idef}
\end{align}
We note that to obtain $I(u,v)$ we have also changed the integration
variables with the corresponding Jacobian
\begin{equation}
  \dd^2 \tilde{z} = \left({\beta \over 2\pi} \right)^2 {\dd^2 z \over \abs{z}^2}\, .
\end{equation}

Finally, we use the conformal Ward identity to reduce the four-point
function $\langle TT {\cal O} {\cal O}\rangle$ to explicit expressions
we can integrate. The conformal Ward identity gives us the following
relation between correlators on the plane (here we assume the
${\cal O}_i$ are conformal primaries):
\begin{equation}
  \langle T(u) {\cal O}_1(z_1) \dotsb {\cal O}_n(z_n) \rangle = \sum_{i=1}^n \left[{h_i \over (u-z_i)^2} + {\partial_{z_i} \over u-z_i} \right]\langle {\cal O}_1(z_1) \dotsb {\cal O}_n(z_n) \rangle\, .
\end{equation}
Since the stress tensor is only a global conformal primary and not a
Virasoro primary, one must be more careful in deriving the Ward
identity with multiple $T$ insertions. Due to the central charge term
in the $TT$ OPE, there is an additional contribution in the Ward
identity. For the case at hand, this result is
\begin{multline}
  \langle T(u) T(v) {\cal O}(z_1) {\cal O}(z_2) \rangle = {c/2 \over (u-v)^4} \langle {\cal O}(z_1) {\cal O}(z_2)\rangle + \left[{2 \over (u-v)^2} + {\partial_v \over u-v} \right]\langle T(v) {\cal O}(z_1) {\cal O}(z_2)\rangle \\
  + \sum_{i=1}^2 \left[{\partial_{z_i}  \over u -z_i} + {h \over (u-z_i)^2} \right]\langle T(v) {\cal O}(z_1) {\cal O}(z_2)\rangle\, .
\end{multline}
Here the first line is the $TT$-OPE contribution to the Ward identity,
while the second line comes from the $T{\cal O}$ OPE applied to the
${\cal O}(z_i)$. Note also that the first term is simply the
disconnected part of the correlation function,
$\langle TT\rangle \langle {\cal O} {\cal O}\rangle$. Using the
explicit form of the three point function
$\langle T {\cal O} {\cal O}\rangle$\footnote{This can be derived
  using the Ward identity and the two point function
  $\langle {\cal O}_1 {\cal O}_2\rangle$, or using the fact that the
  stress tensor is still a global conformal primary and the
  correlator of three global primaries is fixed by (global)
  conformal invariance (in any dimension). Compatibility with the OPE then fixes $C_{T{\cal O} {\cal O}} = h$.},
\begin{align}
  \label{eq:TOO-corr}
  \langle T(u) {\cal O}(z_1) {\cal O}(z_2) \rangle ={}&  {h  \over (u-z_1)^2 (u-z_2)^2} {(z_1-z_2)^2 \over \vert z_1-z_2 \vert^{4h}}\, ,
\end{align}
we can write the connected part of the four point
function as
\begin{align} {\langle T(u) T(v) {\cal O}(z_1) {\cal O}(z_2)
    \rangle^{(c)} \over \langle T(v) {\cal O}(z_1) {\cal O}(z_2)
    \rangle } ={}& {h \over (u - z_1)^2} + {h \over (u - z_2)^2} + {2
                   \over (u-v)^2} \nonumber \\
                 &- {2 \over u-v} \left({(h -1) (u-v) \over (u -
                   z_1)(u -z_2)} +{1 \over u - z_1} + {1 \over u -
                   z_2}\right)\, .
\end{align}
Here we recall that the connected four-point function is defined as
$\langle TT {\cal O} {\cal O}\rangle^{(c)} = \langle TT {\cal O} {\cal
  O}\rangle - \langle TT\rangle \langle {\cal O} {\cal O}\rangle$.

Using this expression for $\langle TT {\cal O} {\cal O}\rangle^{(c)}$
in \eqref{eq:Idef}, the leading correction $G^\text{E}_2$ is therefore given by~\eqref{eq:GE2int} with
\begin{multline}
  I(u,v) = \int {\dd^2 z_1 \dd^2 z_2 \over \abs{z_1 z_2}^{2(1-h)}} {(z_1-z_2)^2 \over (v-z_1)^2 (v-z_2)^2\abs{z_1-z_2}^{4h} } \\
  \times \left[{h \over (u - z_1)^2} + {h \over (u - z_2)^2} + {2
                   \over (u-v)^2} - {2 \over u-v} \left({(h -1) (u-v) \over (u -
                   z_1)(u -z_2)} +{1 \over u - z_1} + {1 \over u -
                   z_2}\right) \right]\, . \label{eq:Ic}
\end{multline}

\section{Evaluating the Green's Function}
\label{sec:integrals}

Having obtained a formal expression~$G_2^{\text{E}}(u,v)$ for the perturbation to the Euclidean Green's function, we must now explicitly evaluate the integral~$I$ and then Fourier transform the result in order to identify the skipped poles.  This explicit evaluation is nontrivial due to the various singularities stemming from the correlators.  Indeed, note that the integrand of~$I$ is integrable near small and large~$|z_i|$ as long as~$h > 0$, and integrable near~$z_1 = z_2$ as long as~$h < 1$; these singularities are therefore not a problem if we restrict to~$h \in (0,1)$ when evaluating the integral (we can then analytically continue in~$h$ to obtain the result for other values of~$h$).

However, the integrand is \textit{not} integrable at~$z_i = u$ and~$z_i = v$ due to the~$1/(u-z_i)^2$ and~$1/(v-z_i)^2$ divergences.  These singularities arise from the~$T\Ocal$ insertions in the connected correlator~$\ev{TT\Ocal\Ocal}^{(c)}$, which presumably must be interpreted in an appropriate distributional way in order to evaluate~$I$.  Our first task, then, is to provide an appropriate distributional interpretation of this object.

\subsection{Homogeneous Distributions}
\label{subsec:homogeneous}

Since the~$T(u)\Ocal(z)$ insertions give divergences of the form~$1/(u-z)^2$, our goal is essentially to extend the complex function~$1/z^2$, which is defined on~$\mathbb{C}\setminus \{0\}$, to a distribution on the entire complex plane.  We shall denote such an extension by~$\Dcal(1/z^2)$.  This distribution is defined by its action on arbitrary test functions~$\varphi(z,\zb)$.  Now, if the region of the complex plane on which~$\varphi$ has support does not include the origin, then we have simply
\be
\label{eq:zn2phi}
\int_{\mathbb{C}} d^2 z \, \varphi(z,\zb) \Dcal\frac{1}{z^2} = \int_{\mathbb{C}} d^2 z \, \frac{\varphi(z,\zb)}{z^2},
\ee
which is convergent because~$\varphi(z,\zb)$ vanishes in a neighborhood of the pole.  However, there are infinitely many ways to extend~$\Dcal(1/z^2)$ to test functions whose support includes the origin (though all such definitions are related by delta functions and derivatives thereof).

To obtain a \textit{unique} extension to the entire complex plane, we make use of the notion of complex homogeneity.  In short, a complex function~$f(z,\zb)$ is said to be complex homogeneous of degree~$(\mu,\nu)$ if for any complex~$\alpha$ we have~$f(\alpha z, \bar{\alpha}\zb) = \alpha^\mu \bar{\alpha}^\nu f(z,\zb)$.  Likewise, an arbitrary complex distribution~$\Dcal f(z,\zb)$ is complex homogeneous of degree~$(\mu,\nu)$ if
\be
\int_{\mathbb{C}} d^2 z \, \varphi\left(\frac{z}{\alpha}, \frac{\zb}{\bar{\alpha}}\right) \Dcal f(z,\zb) = \alpha^{\mu+1} \bar{\alpha}^{\nu+1} \int_{\mathbb{C}} d^2 z \, \varphi(z,\zb) \Dcal f(z,\zb)
\ee
(this follows from a formal change of variables and use of the homogeneity of~$f(z,\zb)$).  Since the complex function~$1/z^2$ is homogeneous of degree~$(-2,0)$ away from the origin, it is natural to require that~$\Dcal(1/z^2)$ be homogeneous of the same degree even when extended to the origin.  As discussed in e.g.~Appendix B of~\cite{GelShi}, for any positive integer~$n$ the homogeneous distribution of degree~$(-n,0)$ is unique (up to an overall multiplicative constant) and given by
\be
\int_{\mathbb{C}} d^2 z \, \varphi(z,\zb) \Dcal^\text{H}\frac{1}{z^n} = \frac{1}{(n-1)!} \int_\mathbb{C} d^2 z \frac{\partial^{n-1}_z \varphi(z,\zb)}{z},
\ee
where the superscript~$\text{H}$ denotes that this is the unique complex homogeneous distribution.  We show in Appendix~\ref{subapp:principalvalue} that~$\Dcal^\text{H}(1/z^2)$ can be written in the principal value-like form
\be
\label{eq:DHprincipalvalue}
\int_{\mathbb{C}} d^2 z \, \varphi(z,\zb) \Dcal^\text{H} \frac{1}{z^2} = \lim_{\eps \to 0} \int_{\mathbb{C} \setminus B_\eps(0)} d^2 z \, \frac{\varphi(z,\zb)}{z^2},
\ee
where~$B_\eps(0)$ is a disk of radius~$\eps$ centered on the origin.  Note that when the region of support of~$\varphi$ does not include the origin, this reproduces~\eqref{eq:zn2phi}.

Given that the homogeneity properties of the correlators leading to~$G_2^\text{E}$ stem from conformal invariance of the CFT, it seems quite natural to impose these homogeneity properties when extending the correlators to distributions on the entire complex plane.  From a more physical perspective, the singularity of a~$T(0)\Ocal(z)$ insertion at~$z = 0$ in the Euclidean section corresponds to a light-cone singularity~$x - t = 0$ in the Lorentzian section (since~$z = x + i \tau = x - t$).  Thus the distributional interpretation in~\eqref{eq:DHprincipalvalue}, which in the Euclidean section just corresponds to excising an infinitesimal disk around any poles, can roughly be interpreted in the Lorentzian section as excising an infinitesimal neighborhood of any light cones connecting operator insertions.

The upshot is that when evaluating~$I$ according to this distributional interpretation of the poles, we may freely rescale complex coordinates, and we may treat any poles by just excising an infinitesimal disk around each one.  (We will leave this distributional understanding implicit to avoid excessive cluttering of notation.)  Evidence for the physical consistency of this approach will be provided in Section~\ref{sec:gravity}, where we compare the final outcome of the CFT calculation to a shock wave analysis in the dual gravitational theory.

\subsection{Evaluating the Euclidean Green's Function}

Having understood how to treat all the poles in~$I(u,v)$, we may now evaluate it explicitly.  Since the prefactor inside the integrand of~\eqref{eq:Ic} is symmetric in~$z_1$ and~$z_2$, we can combine some terms inside the square brackets to obtain
\begin{multline}
I(u,v) = 2h \int_{\mathbb{C}^2} \frac{d^2z_1 \, d^2z_2}{|z_1 z_2|^{2(1-h)}} \, \frac{(z_1-z_2)^2}{(u-z_1)^2 (v-z_2)^2 |z_1-z_2|^{4h}} \\ \times \left[\frac{1}{(u-v)^2} + \frac{h-1}{2} \, \frac{(z_1-z_2)^2}{(v-z_1)^2 (u-z_2)^2}\right],
\end{multline}
which is explicitly symmetric in~$u$ and~$v$.  Then changing to new variables~$(w,w')$ defined by~$z_1 = (u-v) w'(w + 1)/2$,~$z_2 = (u-v) w'(w-1)/2$, one obtains
\begin{multline}
I(u,v) = \frac{2^{7-4h}h}{(u-v)^4} \int_{\mathbb{C}^2} \frac{d^2 w \, d^2 w'}{|w'|^2 |1-w^2|^{2(1-h)}} \, \frac{(w')^2}{((\Sigma-w w')^2-(1-w')^2)^2} \\ \times \left[1+\frac{8(h-1)(w')^2}{((\Sigma-w w')^2-(1+w')^2)^2}\right],
\end{multline}
where we have defined
\be
\label{eq:Sigmadef}
\Sigma \equiv \frac{u+v}{u-v} = \coth(\pi(\tilde{u}-\tilde{v})/\beta) \ .
\ee
Note that in order to reduce~$I$ to this form, we had to make use of the fact that we are treating the poles as homogeneous distributions, allowing us to freely rescale coordinates and pull factors out of the integral.

\begin{figure}[t]
\centering
\includegraphics[width=0.32\textwidth]{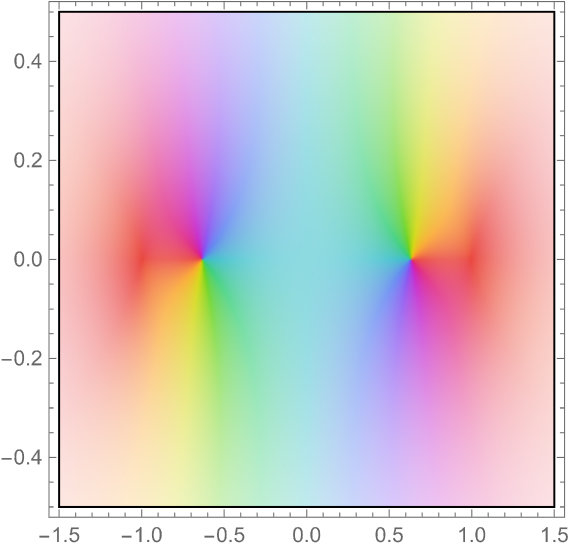}
\includegraphics[width=0.32\textwidth]{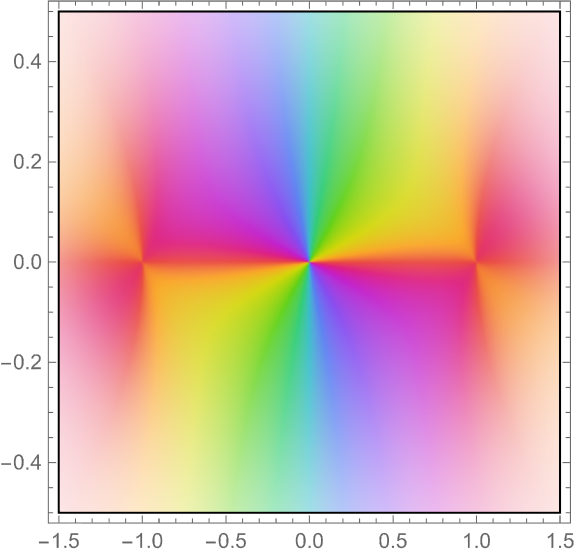}
\includegraphics[width=0.32\textwidth]{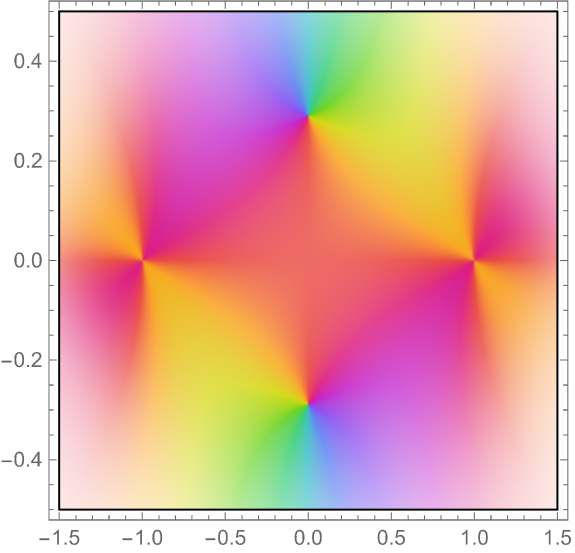}
\caption[]{Plots of~$G_2^\text{E}(\Sigma)$ in the complex~$\Sigma$ plane.  From left to right, the plots correspond to~$h = 0.4$,~$h = h_* \approx 0.74$, and~$h = 0.9$.  $G_2^\text{E}(\Sigma)$ is even in~$\Sigma$, single-valued, vanishes at~$\Sigma = \pm 1$, and diverges at~$\Sigma \to \infty$.  $G_2^\text{E}(\Sigma)$ also exhibits two additional zeros besides~$\Sigma = \pm 1$; for~$h \in (0,h_*)$ these zeros lie in the interval~$\Sigma \in (-1,1)$, while for~$h > h_*$ they are on the imaginary axis.}
\label{fig:G2Eplots}
\end{figure}

We shall relegate most of the technical details of the integration to Appendix~\ref{app:integrals}.  In short, we first split the two integrations by defining
\be
\label{eq:Kdef}
K(w,\Sigma) \equiv \int_\mathbb{C} \frac{d^2 w'}{|w'|^2} \, \frac{(w')^2}{((\Sigma-w w')^2-(1-w')^2)^2} \left[1+\frac{8(h-1)(w')^2}{((\Sigma-w w')^2-(1+w')^2)^2}\right],
\ee
so that
\be
\label{eq:IcKint}
I = \frac{2^{7-4h}h}{(u-v)^4} \int \frac{d^2 w}{|1-w^2|^{2(1-h)}} \, K(w,\Sigma).
\ee
$K(w,\Sigma)$ can be evaluated by using the complex Stokes's theorem to reduce it down to a contour integral, which yields
\begin{multline}
\label{eq:Kfinal}
K(w,\Sigma) = \frac{\pi}{8} \left[-\frac{2(h-1)}{(w+\Sigma)^2} + \frac{(h-1)(1+w\Sigma + (w+\Sigma)^2)}{(w+\Sigma)^3} \ln\left|\frac{(\Sigma+1)(w+1)}{(\Sigma-1)(w-1)}\right|^2 \right. \\ \left. + \frac{(h-1)(1-w\Sigma + (w-\Sigma)^2)+2(1-w \Sigma)}{(w-\Sigma)^3} \ln\left|\frac{(\Sigma-1)(w+1)}{(\Sigma+1)(w-1)}\right|^2\right].
\end{multline}
Inserting this result into~\eqref{eq:IcKint} and applying Stokes' theorem once again reduces~$I$ to a contour integral, which after extensive simplification yields the result~\eqref{eq:Ichatfinal} given in the Appendix.  Using this result in~\eqref{eq:GE2int} finally leaves us with the Euclidean Green's function:
\begin{multline}
\label{eq:G2Eexplicit}
G_2^\text{E}(\tilde{u},\tilde{v}) = \frac{1}{2} \left(\frac{2\pi}{\beta}\right)^{4h} \Bigg\{\frac{\pi^2 \Gamma(1+h)^2 \Gamma(1-2h)}{4h\Gamma(1-h)^2 \Gamma(2h-2)} (1-\Sigma^2) \\
	\times \left[\left(\left(1-(1+2h)\Sigma^2\right)\ln\left|\frac{\Sigma+1}{\Sigma-1}\right|^2 + \frac{2(1+h)}{1-h}\Sigma \right) \frac{(1-\Sigma)^{2(h-1)}}{2^{2h-1}} \pFq{2}{1}{1-h,2(1-h)}{2-h}{\frac{\Sigma+1}{\Sigma-1}} \right. \\
	\left. +  \frac{(1-\Sigma)^{2(h-1)}(1-(1+2h)\Sigma^2)}{2^{2h-1}(h-1)} \pFq{3}{2}{1-h,1-h,2(1-h)}{2-h,2-h}{\frac{\Sigma+1}{\Sigma-1}} - h \Sigma \ln\left|\frac{\Sigma+1}{\Sigma-1}\right|^2 - \frac{1}{h-1} \right] \\
	+ \frac{\pi^2}{2^{2h}} \frac{|1+\Sigma|^{2h} (1-\Sigma)^h}{(1-\overline{\Sigma})^h} \left[2(h^2-1)\Sigma \pFq{2}{1}{h,2h}{1+h}{\frac{\overline{\Sigma}+1}{\overline{\Sigma}-1}} \right. \\ 
	\left. - \frac{(1-h)^2(1-(1+2h)\Sigma^2)}{h} \pFq{3}{2}{h,h,2h}{1+h,1+h}{\frac{\overline{\Sigma}+1}{\overline{\Sigma}-1}}\right]\Bigg\},
\end{multline}
where~$_p F_q$ is a generalized hypergeometric function.  Note that~$G_2^\text{E}$ depends on~$\tilde{u}$ and~$\tilde{v}$ only through the combination~$\Sigma$.  Figure~\ref{fig:G2Eplots} illustrates~$G_2^\text{E}$ in the complex~$\Sigma$ plane for a few values of~$h$.

Although our ultimate goal is to obtain the Fourier transform of the Lorentzian retarded Green's function, it is worth commenting on some features exhibited by~$G_2^\text{E}$ that are not at all obvious from the expression~\eqref{eq:G2Eexplicit}.  All of the following features are expected on physical grounds, and hence serve as checks of our result (including of our treatment of the poles):
\begin{itemize}
	\item From~\eqref{eq:Sigmadef},~$\Sigma$ is translationally invariant on the thermal cylinder.  Thus the fact that~$G_2^\text{E}$ is only a function of~$\Sigma$ ensures that it too is translationally invariant.  It is also even in~$\Sigma$, as can be seen graphically or checked numerically.  This evenness corresponds to the exchange symmetry~$\tilde{u} \leftrightarrow \tilde{v}$.
	\item The generalized hypergeometric function~$_pF_q(z)$ is multi-valued, with its branch cut typically taken to be on the half-line~$z > 1$, so the hypergeometric functions and the radicals in~$G_2^\text{E}$ individually have branch cuts along the half-line~$\Sigma > 1$.  However, these cuts cancel out: as can be seen in Figure~\ref{fig:G2Eplots},~$G_2^\text{E}$ is in fact single-valued in~$\Sigma$.
	\item $G_2^\text{E}$ vanishes at~$\Sigma = \pm 1$, which on the thermal cylinder corresponds to~$\Re(\tilde{u} - \tilde{v}) \to \pm \infty$: in other words,~$G_2^\text{E}$ vanishes at large spatial separation.  Conversely,~$G_2^\text{E}$ diverges at large~$\Sigma$, which on the thermal cylinder corresponds to the coincidence limit~$\tilde{u} \to \tilde{v}$.
	\item When~$h = 1$,~$G_2^\text{E}$ vanishes entirely.  This behavior is not at all obvious from the integral form~\eqref{eq:GE2int}, but it is physically expected:~$h = 1$ corresponds to a marginal deformation of the CFT, which does not break conformal invariance.  The Green's function should therefore be unchanged under such a deformation, consistent with the behavior we have found.
\end{itemize}
Besides the zeros at~$\Sigma = \pm 1$, $G_2^\text{E}$ also exhibits two additional zeros.  When~$h$ is sufficiently small, these zeros appear in the interval~$\Sigma \in (-1,1)$; for larger~$h$, they move onto the imaginary axis.  The threshold value at which they transition between these behaviors is~$h_* \approx 0.739837$, shown in the middle plot in Figure~\ref{fig:G2Eplots}.

\subsection{Fourier Transforming}
\label{subsec:Fourier}

In order to identify the perturbation to the skipped poles, we must now compute the Fourier transform of the perturbation to the Lorentzian retarded two-point function.  We will not be able to obtain a closed-form expression for the Fourier transform for all~$h$.  We will therefore set up the calculation in general, but then evaluate the Fourier transform only for certain tractable values of~$h$.

\begin{figure}[t]
\centering
\subfloat[]{
\includegraphics[page=1]{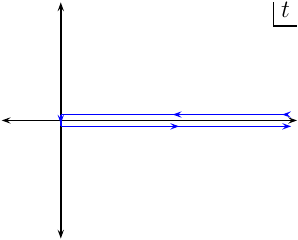}
\label{subfig:tFouriercontour}
}
\hspace{2cm}
\subfloat[]{
\includegraphics[page=2]{Figures/Paper-pics.pdf}
\label{subfig:tmodifiedcontour}
}
\caption[]{\subref{subfig:tFouriercontour}: the contour of integration in~$t$ for evaluating the Fourier transform~\eqref{eq:tFouriertransform}.  \subref{subfig:tmodifiedcontour}: the contour of integration can be deformed into a keyhole contour that runs along the branch cut~$t > |x|$ and circles the branch point~$t = |x|$.  The branch point~$t = x$ is singular, so when~$x > 0$ we must be careful to keep track of the singular contributions from the circle of radius~$\eps$ around the branch point and from the integral along the cut from~$t = x + \eps$ to infinity.  These contributions are individually singular as~$\eps \to 0$, but the divergences must cancel out.}
\label{fig:tFourier}
\end{figure}

Note that due to the translational invariance of~$G_2^\text{E}(\tilde{u},\tilde{v})$ on the thermal cylinder, it will be convenient to work in terms of~$\Sigma$; that is, we write~$G_2^\text{E}(\Sigma,\overline{\Sigma})$ as a function of~$\Sigma$ and~$\overline{\Sigma}$.  In terms of Lorentzian coordinates~$(t,x)$,~$\Sigma$ and~$\overline{\Sigma}$ are light-cone coordinates:
\be
\label{eq:SigmaLorentzian}
\Sigma(t,x) = \coth(\pi(x-t)/\beta), \qquad \overline{\Sigma}(t,x) = \coth(\pi(x+t)/\beta).
\ee
The Lorentzian-signature retarded two-point function is then obtained from analytically continuing the Euclidean two-point function:
\be
\label{eq:DefLorGF}
G_2(t,x) = -i\Theta(t) \left[G_2^\text{E}\left(\Sigma(t-i\eps,x),\overline{\Sigma}(t-i\eps,x)\right)-G_2^\text{E}\left(\Sigma(t+i\eps,x),\overline{\Sigma}(t+i\eps,x)\right) \right],
\ee
with~$\eps$ an infinitesimal positive real number.  The Lorentzian Fourier transform can thus be computed by an appropriate contour integral over the Euclidean two-point function (see e.g.~\cite{Ramirez:2020qer}):
\be
\label{eq:tFouriertransform}
G_2(\omega,x) = -i \int_C dt \, e^{i\omega t} G_2^\text{E}\left(\Sigma(t,x),\overline{\Sigma}(t,x)\right),
\ee
where the contour~$C$ runs around the positive~$t$ axis, as shown in Figure~\ref{subfig:tFouriercontour}.

In order to evaluate~\eqref{eq:tFouriertransform}, we must therefore understand the analytic structure of~$G_2^\text{E}(\Sigma,\overline{\Sigma})$ when~$\Sigma$ and~$\overline{\Sigma}$ are treated as independent variables.  First,~$G_2^\text{E}$ is singular when~$\Sigma \to \infty$, corresponding to~$t = x$ (there is no analogous singularity when~$\overline{\Sigma} \to \infty$).  So for~$x > 0$, the contour of integration shown in~\ref{subfig:tFouriercontour} runs around a singular point.  More importantly, the radicals and hypergeometric functions in~\eqref{eq:G2Eexplicit} exhibit branch cuts whenever~$\Sigma \in (-\infty,-1)$,~$\Sigma \in (1,\infty)$,~$\overline{\Sigma} \in (-1,\infty)$, or~$\overline{\Sigma} \in (1,\infty)$.  Na\"ively, from~\eqref{eq:SigmaLorentzian} these suggest that~$G_2^\text{E}$ may have a branch cut along the entire real~$t$ axis.  However, in Appendix~\ref{subapp:tdisc} we perform a careful computation of the discontinuity of~$G_2^\text{E}$ across the real~$t$ axis, and we find that the cuts cancel out in the interval~$t \in (-|x|,|x|)$.  That is,~$G_2^\text{E}$ only exhibits branch cuts across the portions~$|t| > |x|$.

As a brief aside, the analytic behavior of~$G_2^\text{E}(\Sigma(t,x),\overline{\Sigma}(t,x))$ we just identified gives rise to the correct behavior expected of the real-time Lorentzian two-point function~$G_2(t,x)$.  Combining~\eqref{eq:DefLorGF} with the discontinuity~\eqref{eq:DeltaG2} calculated in the Appendix, in Figure~\ref{fig:GFplot} we plot~$G_2(t,x)$  for the case~$h = 1/2$ (other values of~$h$ exhibit the same qualitative behavior).  The fact that~$G_2^\text{E}$ only has a branch cut when~$|t| > |x|$ implies that~$G_2$ has support inside the future light cone but is zero outside, as expected for a CFT perturbed away from its conformal point\footnote{There is also a distributional contribution with support \textit{on} the right light cone; this contribution will be important when Fourier transforming, but we do not illustrate it in Figure~\ref{fig:GFplot}.}.  Its right-moving character comes from our choice to analyze only the holomorphic component of the stress tensor, as discussed at the end of Section~\ref{sec:PertExp}.  These behaviors serve as further physical checks of our results, and also instantiate a universal mechanism for thermalization dynamics in $(1+1)$-dimensional quantum field theories -- see~\cite{Davison:2024msq}, especially their Figure~1. 

\begin{figure}[t]
\centering
\includegraphics[width=0.8\textwidth]{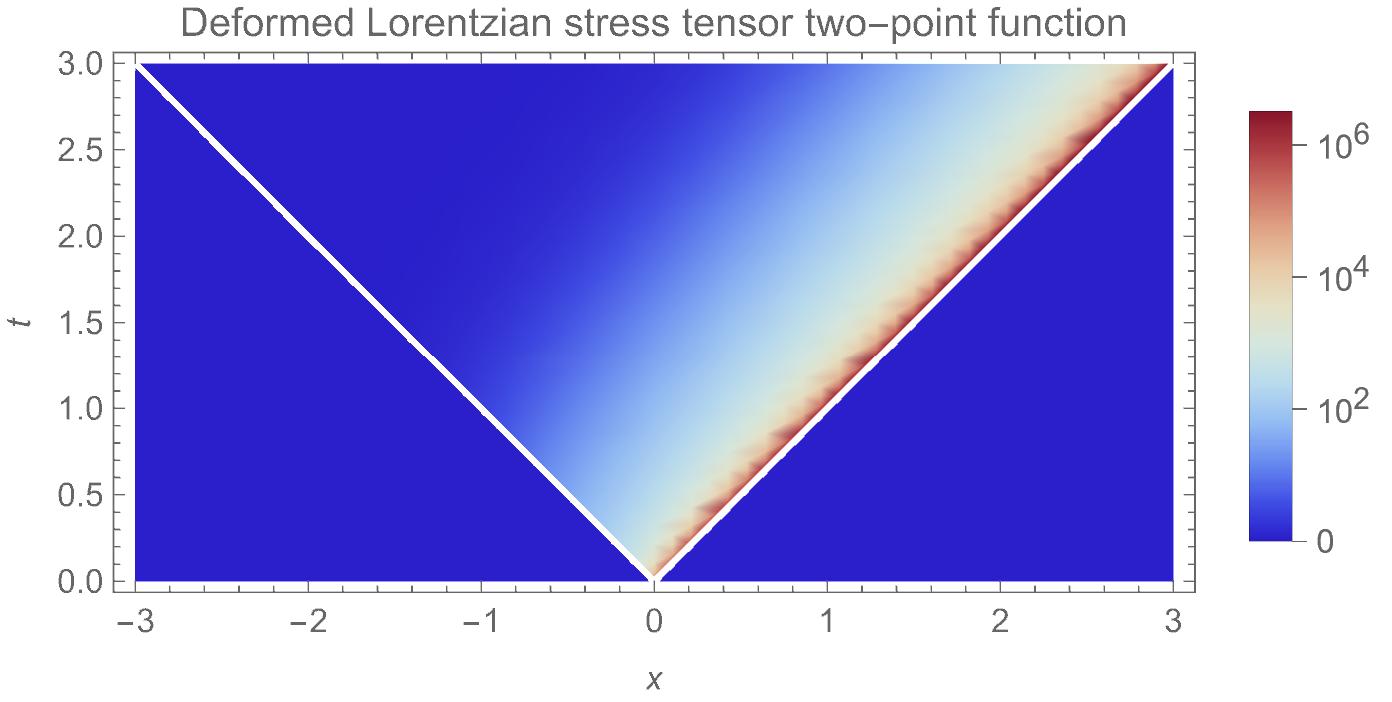}
\caption[]{The~$O(\lambda^2)$-contribution to the deformed Lorentzian two-point function of the holomorphic component of the stress tensor,~$G_2(t,x)$; c.f.~equations~\eqref{eq:DefLorGF} and~\eqref{eq:DeltaG2}.  Here we take~$h= 1/2, \beta = 1$.  It is zero everywhere outside the light cone, as expected; it is nonzero everywhere inside the future light cone, approaching zero at the left light cone and diverging at the right light cone.}
\label{fig:GFplot}
\end{figure}

Turning back to the evaluation of the Fourier transform in~$t$, the fact that~$G_2^\text{E}$ only exhibit a branch cut for~$|t| > |x|$ allows us to deform the contour of integration into the keyhole contour shown in Figure~\ref{subfig:tmodifiedcontour}.  The Fourier transform can thus be decomposed into two contributions: one from the discontinuity across the branch cut~$t > |x|$, and another from the circle around the singularity at~$t = x$ which only contributes when~$x > 0$:
\begin{subequations}
\begin{align}
F_\mathrm{pole}(\omega,x) &= -i \Theta(x) \int_{C_\eps(x)} dt \, e^{i\omega t} G_2^\text{E}(\Sigma(t,x),\overline{\Sigma}(t,x)), \label{subeq:Fpole} \\
F_\mathrm{cut}(\omega,x) &= -i \int_{|x|+\eps}^\infty dt \, e^{i\omega t} \Delta G_2(t,x), \label{subeq:Fcut}
\end{align}
\end{subequations}
where~$\Delta G_2$ is the discontinuity of~$G_2^\text{E}$ across the branch cut and~$\eps$ is an infinitesimal positive quantity to be taken to zero; it is only needed when~$x > 0$.  The full Fourier transform for identifying the skipped poles is then just
\be
\label{eq:xFourier}
G_2(\omega,k) = \int_{-\infty}^\infty dx \, e^{-ikx} \left[F_\mathrm{pole}(\omega,x) + F_\mathrm{cut}(\omega,x)\right].
\ee

Now,~$F_\mathrm{pole}$ can be evaluated for general~$h$; the result is given in~\eqref{eq:Fpolegeneral}.  On the other hand, while it is straightforward to compute the discontinuity~$\Delta G_2(t,x)$ (given in~\eqref{eq:DeltaG2}) that appears in~$F_\mathrm{cut}$, evaluating its integral for general~$h$ is more challenging.  To proceed, we will therefore now restrict to specific values of~$h$ that make the calculations tractable.  In what follows, we will work with dimensionless variables~$\that = \pi t/\beta$,~$\xhat = \pi x/\beta$,~$\omegahat = \beta \omega/\pi$, and~$\khat = \beta k/\pi$.

\subsubsection*{The case $h = 1/2$}

The case~$h = 1/2$ is tractable since the discontinuity across the cut simplifies substantially.  Using~\eqref{eq:DeltaG2}, we find for~$h = 1/2$ that the contribution from the branch cut reduces to
\be
F_\mathrm{cut}(\omegahat,\xhat) = -\frac{\pi^4}{4\beta} \int_{|\xhat|+\eps}^\infty d\that \, e^{i\hat{\omega}\that} \csch^3(\that-\xhat) \left[3 \sinh(2(\that-\xhat)) + (\that+\xhat)\left(3+\cosh(2(\that-\xhat))\right)\right].
\ee
When~$\xhat < 0$, there is no singularity at the lower limit of integration, and we can perform the integral over~$\that \in (-\xhat,\infty)$ without worrying about the cutoff~$\eps$.  On the other hand, when~$\xhat > 0$ we evaluate the integral for nonzero~$\eps$ and then expand around~$\eps = 0$.  The result is
\begin{multline}
\label{eq:Fcuthhalf}
F_\mathrm{cut}(\omegahat,\xhat) = -\frac{\pi^4}{2\beta} \Theta(-\xhat)\left\{3 e^{i\hat{\omega} \xhat} \left[B\left(e^{4\xhat};\frac{1-i\hat{\omega}}{2},-2\right) - B\left(e^{4\xhat};\frac{5-i\hat{\omega}}{2},-2\right)\right] \right. \\ + 2 e^{-i\hat{\omega} \xhat} \left[ \frac{e^{2\xhat}}{(1-i\hat{\omega})^2} \pFq{3}{2}{3,\frac{1-i\hat{\omega}}{2},\frac{1-i\hat{\omega}}{2}}{\frac{3-i\hat{\omega}}{2},\frac{3-i\hat{\omega}}{2}}{e^{4\xhat}} + \frac{6e^{6\xhat}}{(3-i\hat{\omega})^2} \pFq{3}{2}{3,\frac{3-i\hat{\omega}}{2},\frac{3-i\hat{\omega}}{2}}{\frac{5-i\hat{\omega}}{2},\frac{5-i\hat{\omega}}{2}}{e^{4\xhat}} \right. \\ \left.\left. + \frac{e^{10\xhat}}{(5-i\hat{\omega})^2} \pFq{3}{2}{3,\frac{5-i\hat{\omega}}{2},\frac{5-i\hat{\omega}}{2}}{\frac{7-i\hat{\omega}}{2},\frac{7-i\hat{\omega}}{2}}{e^{4\xhat}}\right]\right\} -\frac{\pi^4}{\beta} \Theta(\xhat) e^{i\hat{\omega} \xhat}\left\{ \frac{\xhat}{\eps^2} + \frac{5 + 4i\hat{\omega}\xhat}{2\eps} \right. \\ -\frac{i\omegahat}{2}(5+2i\omegahat\xhat)\ln(2\eps) + \frac{1}{12} \left[-6i\hat{\omega}\left(5 + 2 i \hat{\omega} \xhat\right) H_{-\frac{1+i\hat{\omega}}{2}} +36 i \hat{\omega} + 2\xhat \left(1 -9\hat{\omega}^2\right) \right. \\ \left.\left. - 3 \hat{\omega}^2 \psi^{(1)}\left(\frac{1-i\hat{\omega}}{2}\right)\right] + O(\eps)\right\},
\end{multline}
where~$B(z;a,b)$ is the incomplete beta function,~$H_z$ is the generalized harmonic number, and~$\psi^{(1)}(z)$ is the polygamma function.

The contribution from the pole when~$\xhat > 0$ cannot be obtained from the general result~\eqref{eq:Fpolegeneral} because~\eqref{eq:Fpolegeneral} is only valid for~$h \in (0,1/2)$.  Instead, we compute the pole contribution from scratch by first taking the~$h \to 1/2$ limit of~$G_2^\text{E}(\Sigma(t,x),\overline{\Sigma}(t,x))$ to obtain the Green's function at~$h = 1/2$.  We then set~$\that = \xhat + \eps e^{i\theta}$ and expand in~$\eps$, keeping careful track of branch cuts.  We finally integrate over~$\theta \in(0,2\pi)$ (which avoids the branch cut at~$\theta = 0$) to eventually obtain
\begin{multline}
\label{eq:Fpolehhalf}
F_\mathrm{pole}(\omegahat,\xhat) = \frac{\pi^4}{\beta} \Theta(\xhat) e^{i\hat{\omega} \xhat}\left\{\frac{\xhat}{\eps^2} + \frac{5 + 4i\hat{\omega}\xhat}{2\eps} - \frac{i\omegahat}{2} (5+2i\omegahat \xhat) \ln \left(\frac{\eps}{2}\right) \right. \\ - \frac{1}{24} \left[6\hat{\omega}^2\left(\mathrm{Li}_2(e^{-4\xhat}) - 4 \, \mathrm{Li}_2(e^{-2\xhat})\right) + 24 \csch(\xhat) \sech(\xhat) - 60 i \hat{\omega} \ln(\coth(\xhat)) \phantom{\frac{1}{2}}\right. \\ \left.\left. \phantom{\frac{1}{2}} - 4\xhat\left(1 + 3\hat{\omega}^2\right) + 3\pi^2 \hat{\omega}^2\right] + O(\eps)\right\},
\end{multline}
where~$\mathrm{Li}_2(z)$ is the dilogarithm.  Comparing~\eqref{eq:Fcuthhalf} and~\eqref{eq:Fpolehhalf}, note that the divergent terms in~$\eps$ cancel, so we can safely take~$\eps \to 0$ once we combine the contributions from the pole and the cut.

Now we just need the final Fourier transform in~$x$~\eqref{eq:xFourier}, which requires us to come up with a distributional interpretation of~$F_\mathrm{cut} + F_\mathrm{pole}$.  Specifically, at large and small~$|\xhat|$, we find that
\be
F_\mathrm{pole}(\omegahat,\xhat) + F_\mathrm{cut}(\omegahat,\xhat) = \begin{cases} O(e^{2\xhat}), & \xhat \to -\infty, \\ - \frac{\pi^4}{\beta|\xhat|} + O(\ln|\xhat|), & |\xhat| \ll 1, \\ O(e^{i\omegahat\xhat} \xhat), & \xhat \to \infty. \end{cases}
\ee
So the Fourier transform poses no problem at large negative~$x$.  At large positive~$x$, the divergent behavior can be regulated in the standard way by giving~$\omega$ an infinitesimal positive imaginary part.  It is the small-$x$ behavior~$1/|x|$ that needs to be given a distributional interpretation.  In Appendix~\ref{subapp:absvalue} we review an extension of~$1/|x|$ to a distribution~$\Dcal(1/|x|)$ on the real line; its action on a test function is given by (see~\eqref{subeq:oneoverabsx})
\be
\label{eq:oneoverabsx}
\int_{-\infty}^\infty dx \, \varphi(x) \Dcal \frac{1}{|x|} = \lim_{\eps \to 0} \left[\int_{|x| \geq \eps} \frac{\varphi(x)}{|x|} \, dx + 2 \varphi(0) \ln \eps \right].
\ee
Using this interpretation, the final Fourier transform in~$x$ can be expressed as
\be
G_2(\omegahat,\khat) = \frac{\beta}{\pi} \lim_{\eps \to 0} \left[\int_{|\xhat| \geq \eps} d\xhat \, e^{-i\hat{k}\xhat} \left(F_\mathrm{pole}(\omegahat,\xhat) + F_\mathrm{cut}(\omegahat,\xhat)\right) - \frac{2\pi^4}{\beta} \ln \eps\right].
\ee
The integral is now convergent (assuming we give~$\omega$ an infinitesimal positive imaginary part), so it can be evaluated and the~$\eps \to 0$ limit can be taken.  The result is
\begin{multline}
\label{eq:G2hhalf}
G_2(\omegahat,\khat) = \frac{\pi^3}{2(\hat{\omega}-\hat{k})^2} \left[2\hat{\omega}(\hat{\omega}-3\hat{k}) \phantom{H_{\frac{1}{2}}} \right. \\ \left. - (\hat{\omega}-2\hat{k})(\hat{k}+\hat{\omega})\left(H_{-\frac{1}{2}-\frac{i(\hat{\omega}-\hat{k})}{4}} + H_{-\frac{1}{2}-\frac{i(\hat{\omega}+\hat{k})}{4}} + \ln 16\right) \right].
\end{multline}

\paragraph{Comment: Non-uniqueness of~$\Dcal(1/|x|)$.}  Note that the extension of~$1/|x|$ to the distribution~$\Dcal(1/|x|)$ is not unique.  In analogy with our unique identification of~$\Dcal^\text{H}(1/z^2)$ in Section~\ref{subsec:homogeneous}, one might hope that a unique extension of~$1/|x|$ to a distribution on the entire real line could be found by enforcing homogeneity of~$\Dcal(1/|x|)$.  This is not possible: from~\eqref{eq:oneoverabsx}, we have that for any~$\alpha > 0$,
\be
\int_{-\infty}^\infty dx \, \varphi(x/\alpha) \Dcal \frac{1}{|x|} = \int_{-\infty}^\infty dx \, \varphi(x) \Dcal \frac{1}{|x|} + 2\varphi(0) \ln \alpha,
\ee
which indicates that~$\Dcal(1/|x|)$ is a so-called associated homogeneous distribution of degree~$-1$:
\be
\Dcal\frac{1}{|\alpha x|} = \alpha^{-1}\left(\Dcal \frac{1}{|x|} + 2 \ln \alpha \, \delta(x)\right).
\ee
Any other extension of~$1/|x|$ to a distribution on the real line is related to~$\Dcal(1/|x|)$ by delta functions and derivatives thereof.  Since~$\delta^{(n)}(x)$ is a homogeneous distribution of degree~$-(n+1)$, such delta functions cannot modify~$\Dcal(1/|x|)$ into a homogeneous distribution.  Thus no homogeneous extension of~$1/|x|$ to the entire real line exists.  At best, we can demand that~$1/|x|$ be extended to an \textit{associated} homogeneous distribution of degree~$-1$, which means that it is only ambiguous up to an additive delta function~$\delta(x)$ (and no derivatives thereof).

The non-uniqueness of the extension of~$1/|x|$ to the real line leads to non-uniqueness in the Fourier transform~$G_2(\omegahat,\khat)$; this non-uniqueness boils down to an additive~$\khat$-independent constant.  For example, if we were to interpret~$1/|x|$ as~$\Dcal(1/|x|) + a \delta(x)$ for some~$a$, then the Green's function would transform as
\be
\label{eq:G2ambiguity}
G_2(\omegahat,\khat) \to G_2(\omegahat,\khat) - \pi^3 a.
\ee
Since such an additive constant is nonsingular, it will not affect the location of the skipped poles, so this ambiguity will not affect our analysis. We will revisit the effect of such contact terms in Section~\ref{subsec:poles}.

\subsubsection*{The case $h$ near $1$}

The other tractable special case we will examine is the behavior near~$h = 1$.  When~$h = 1$, the perturbation is marginal and~$G_2^\text{E}$ vanishes, as expected, but we can work to next-to-leading order in~$(h-1)$:
\be
G_2^\text{E}(t,x) = \frac{\pi^2}{8} \left(\frac{2\pi}{\beta}\right)^{4h} \csch^4(x-t) \left[1+ 2 \csch(t+x) \sinh(3t-x)\right](h-1) + O(h-1)^2.
\ee
The piece linear in~$(h-1)$ only has poles at~$t = \pm x$ but no branch cut (as could also be inferred by expanding the discontinuity~$\Delta G_2$ given in~\eqref{eq:DeltaG2})\footnote{The pole at~$t = -x$ is not present for any \textit{finite}~$(h-1)$; it appears due to the fact that we are working perturbatively in~$(h-1)$.}.  Thus the contour integral for the Fourier transform in~$t$ only picks up the residue from whichever pole is on the positive~$t$ axis, giving
\begin{multline}
F_\mathrm{pole}(\omegahat,\xhat) = -\pi^3(h-1) \left(\frac{2\pi}{\beta}\right)^{4h-1} e^{i\hat{\omega}|\xhat|} \left\{ \Theta(-\xhat) \coth(2\xhat)\csch^2(2\xhat) \phantom{\frac{1}{2}} \right. \\ \left. + 
\frac{1}{8} \Theta(\xhat) \left[i\hat{\omega}(\hat{\omega}^2 - 12) + 4\coth(2\xhat)(2+\hat{\omega}^2 + 2i\hat{\omega} \coth(2\xhat) - 2 \coth^2(2\xhat)\right]\right\} \\ + O(h-1)^2.
\end{multline}
At large and small~$|\xhat|$,
\be
F_\mathrm{pole}(\omegahat,\xhat) = \pi^3 (h-1) \left(\frac{2\pi}{\beta}\right)^{4h-1} \begin{cases} O(e^{4\xhat}), & \xhat \to -\infty, \\ \frac{1}{8|\xhat|^3} e^{-i\hat{\omega}\xhat} + O(\xhat^0), & |\xhat| \ll 1, \\ O(e^{i\omegahat\xhat}), & \xhat \to \infty, \end{cases}
\ee
so again the Fourier transform in~$x$ does not pose any problems at large~$|\xhat|$ (provided we give~$\omegahat$ a small positive imaginary part), but needs to be interpreted distributionally at small~$|\xhat|$ due to the~$1/|x|^3$ divergence.  Here we will use the distribution~$\widetilde{\Dcal}(1/|x|^3)$ given in~\eqref{subeq:Dtildex3} to write
\be
G_2(\omegahat,\khat) = \frac{\beta}{\pi} \lim_{\eps \to 0} \left[\int_{|\hat{x}| \geq \eps} e^{-i\khat\xhat} F_\mathrm{pole}(\omegahat,\xhat) d\xhat - \frac{\pi^3(h-1)}{8} \left(\frac{2\pi}{\beta}\right)^{4h-1} \left(\frac{1}{\eps^2} + (\omegahat + \khat)^2 \ln \eps \right)\right].
\ee
We are eventually left with
\begin{multline}
\label{eq:G2hone}
G_2(\omegahat,\khat) = \frac{\pi^3(h-1)}{24} \left(\frac{2\pi}{\beta}\right)^{4h-2} \left[\frac{8\hat{k}+\hat{\omega}(3\hat{k}^2-15\hat{k}\hat{\omega}-32)+9(\hat{\omega}^3+\hat{k}^3) + 12i\hat{\omega}(\hat{\omega}+\hat{k})}{\hat{\omega}-\hat{k}} \right. \\ \left. \phantom{\frac{1}{2}} + 3(\hat{\omega}+\hat{k})^2\left(H_{-1-\frac{i(\hat{\omega}-\hat{k})}{4}} + H_{-1-\frac{i(\hat{\omega}+\hat{k})}{4}} + \ln 16\right)\right] + O(h-1)^2.
\end{multline}

\subsection{Skipped Poles}
\label{subsec:poles}

Having obtained the frequency-space retarded Green's function (at least for a couple specific values of~$h$), we can now investigate the perturbation of the skipped poles.  The full Green's function is
\be
G(\omegahat,\khat) = G_0(\omegahat,\khat) + \lambda^2 G_2(\omegahat,\khat) + O(\lambda^3),
\label{eq:GL-full}
\ee
where the unperturbed Green's function, in dimensionless variables, is \cite{Haehl:2018izb, Ramirez:2020qer} 
\be
\label{eq:G0}
G_0(\omegahat,\khat) = \frac{\pi^3 c}{6\beta^2} \, \frac{\hat{\omega}(\hat{\omega}^2+4)}{\hat{\omega}-\hat{k}}.
\ee
This result follows from the prescription for Fourier transforming described in Section~\ref{subsec:Fourier}. The attentive reader may note that it differs from \eqref{eq:GR0-WI}, which we obtained by solving the Ward identities. One can check that \eqref{eq:G0} and \eqref{eq:GR0-WI} differ by a polynomial in~$\omega$ and~$k$, i.e.~they differ by contact terms. In the remainder of this subsection, we will first use the expansion \eqref{eq:GL-full}, with \eqref{eq:G0}, to obtain the perturbed pole skipping points using the procedure described in Section~\ref{sec:Ward}, and then we will return to discuss the effect of possible contact term ambiguities on the analysis.

\subsubsection*{The case $h = 1/2$}

The expression for~$G_2$ when~$h = 1/2$ is given in equation~\eqref{eq:G2hhalf}.  Requiring that~$G(\omegahat_*,\khat_*)$ vanish to order~$\lambda^2$ imposes the conditions~$\omega_0 = 2i$ and
\be
\label{eq:omega2hhalf}
\omega_2 = \frac{3\beta^2}{8c(k_0-2i)}\left[8+12ik_0-2(2+ik_0+k_0^2)\left(H_{-\frac{ik_0}{4}} + H_{\frac{ik_0}{4}} + \ln 16\right)\right],
\ee
which locates for us the surface of zeros of~$G$.  Next, requiring that~$1/G(\omegahat_*,\khat_*)$ vanish to order~$\lambda^2$ imposes~$k_0 = \omega_0$ and
\be
\label{eq:k2hhalf}
k_2 = \omega_2 - \frac{6\beta^2\omega_0}{c(\omega_0^2 + 4)}\left[H_{-\frac{1+i\omega_0}{2}} + \ln 4 -2\right],
\ee
giving us the singular surface of~$G$.  To find the intersection of these surfaces, we take the limit~$k_0 \to \omega_0 = 2i$ in~\eqref{eq:omega2hhalf} and the limit~$\omega_0 \to 2i$ in~\eqref{eq:k2hhalf} to find that
\be
\omega_2 = 0, \qquad k_2 = \frac{3(\pi^2-8)\beta^2 i}{4c}.
\ee
Reinserting the factors of~$\pi/\beta$, we have thus found that the perturbed skipped pole is located at
\be
\label{eq:omega-k-hhalf}
h = \frac{1}{2}: \quad (\omega_*,k_*) = \frac{2\pi i}{\beta} \left(1 + O(\lambda^3), 1 + \frac{3 (\pi^2-8) \beta^2}{8c} \lambda^2 + O(\lambda^3)\right),
\ee
and its pole-skipping velocity is
\be
\label{eq:vPShhalf}
h = \frac{1}{2}: \quad v_\mathrm{PS} = 1 - \frac{3(\pi^2-8)}{8} \frac{\beta^2 \lambda^2}{c} + O(\lambda^3).
\ee

\subsubsection*{The case $h$ near $1$}

We can perform the same procedure using the expression~\eqref{eq:G2hone} for~$G_2$ near~$h = 1$.  Expanding the~$\omega_i$ and~$k_i$ in powers of~$(h-1)$, we find that requiring~$G$ to vanish imposes~$\omega_0 = 2i$ and
\begin{multline}
\label{eq:omega2hone}
\omega_2 = \frac{\pi^2(h-1)}{8c} \left(\frac{2\pi}{\beta}\right)^{4h-4} (k_0 + 2i) \left[44 - 24 i k_0 + 9k_0^2 \phantom{H_{\frac{1}{2}}} \right. \\ \left. - 3(k_0 - 2i)^2\left(H_{-\frac{6+ik_0}{4}} + H_{-\frac{6-ik_0}{4}} - \ln 16\right)\right] + O(h-1)^2.
\end{multline}
Next, requiring~$1/G(\omegahat_*,\khat_*)$ to vanish gives simply
\be
k_2 = \omega_2 + O(h-1)^2.
\ee
Taking the limit~$k_0 \to \omega_0 = 2i$ in~\eqref{eq:omega2hone} gives~$\omega_2 = 0$, and hence the skipped poles are undisturbed to~$O(\lambda^2(h-1)^2)$:
\be
\label{eq:omega-k-h1}
h \mbox{ near } 1: \quad (\omega_*,k_*) = \frac{2\pi i}{\beta}\bigg(1 + O(\lambda^2 (h-1)^2,\lambda^3), 1 + O(\lambda^2 (h-1)^2,\lambda^3) \bigg).
\ee
The pole-skipping velocity is therefore unperturbed to this order:
\be
\label{eq:vPShone}
h \mbox{ near } 1: \quad v_\mathrm{PS} = 1 + O(\lambda^2 (h-1)^2,\lambda^3).
\ee

\subsubsection*{Role of contact terms}
Finally, we return to the question of contact terms. We have so far seen at least two places where such terms arise. First, there is the ambiguity in the Fourier transform \eqref{eq:G2ambiguity}, and then the discrepancy between~\eqref{eq:G0} and~\eqref{eq:GR0-WI}. In both cases, the ambiguities are polynomials in $\omega$ and $k$, i.e.~they are contact terms. These may hint that our regularization scheme or the distributional nature of the Fourier transforms involved don't respect the Ward identities at coincident points, or perhaps, in the case of the disagreement between~\eqref{eq:G0} and~\eqref{eq:GR0-WI}, that the definition of the stress tensor may be different in the two calculations (as mentioned in footnote $4$ of \cite{Davison:2024msq} and section $2$ of \cite{Romatschke:2009ng})\footnote{We are grateful to Richard Davison for discussions on these points.}.

However, we note that such non-singular additive contributions don't affect the location of the perturbed skipped poles. To see why, imagine writing~$G$ as
\be
G(\omega,k) = \frac{P(\omega,k)}{Q(\omega,k)},
\ee
where~$P$ and~$Q$ are nonsingular functions of~$(\omega,k)$ in~$\mathbb{C}^2$.  Then the zero surface corresponds to~$P(\omega,k) = 0$ and the singular surface to~$Q(\omega,k) = 0$, and the skipped poles lie at the intersection of these surfaces.  Now consider modifying~$G$ by an additive contribution of~$N(\omega,k)$:
\be
G(\omega,k) = \frac{P(\omega,k)}{Q(\omega,k)} + N(\omega,k) = \frac{P(\omega,k)+Q(\omega,k)N(\omega,k)}{Q(\omega,k)}.
\ee
The singular surface is still determined by~$Q(\omega,k) = 0$, but the zero surface has been modified to~$P(\omega,k) + Q(\omega,k)N(\omega,k) = 0$.  The \textit{intersection} of the two surfaces is now determined by a simultaneous solution of~$Q(\omega,k) = 0$ and~$P(\omega,k) + Q(\omega,k)N(\omega,k) = 0$.  As long as~$N(\omega,k)$ is not singular where~$Q(\omega,k)$ vanishes, these equations amount to the original intersection equations~$P(\omega,k) = 0$ and~$Q(\omega,k) = 0$; in other words, the additive contribution~$N(\omega,k)$ had no effect on the skipped poles. The only way for~$N(\omega,k)$ to affect the skipped poles is for~$N(\omega,k) Q(\omega,k)$ to be nonzero where~$Q(\omega,k)$ vanishes.  The possible ambiguities we identified above both lead to corrections $N(\omega,k)Q(\omega,k)$ to the zero surface that vanish on the singular surface $Q(\omega,k) =0$, so that the skipped poles are unaffected (at least to the order at which we are working).

\section{Holographic Butterfly Velocity}
\label{sec:gravity}

While we have computed the locations of the skipped poles using purely CFT techniques, in Section~\ref{subsec:homogeneous} we made use of a particular distributional interpretation of the singular integrals that arise in CFT perturbation theory.  We now wish to provide a verification that our distributional interpretation is physically correct.  To do so, we rely on the observation that to order~$\lambda^2$, the perturbed Green's function computed in Section~\ref{sec:integrals} must agree with that of a deformed holographic CFT.  Since in a holographic theory the skipped poles match the Lyapunov exponent and butterfly velocity~\cite{Chua:2025vig}, we can check our CFT results by performing a holographic calculation of the butterfly velocity of the deformed theory and verifying that it matches the pole-skipping velocity found in the previous section.  Gravitationally, this means computing the propagation of a shock wave on a Ba\~{n}ados-Teitelboim-Zanelli BTZ black hole background \cite{Banados:1992wn}, perturbed by turning on a massive scalar field.

We will proceed in two steps.  First, we turn on a static and planar-symmetric massive scalar field on the planar BTZ black hole in a probe approximation, paying careful attention to ensure that the normalization of the bulk solution is chosen to match the CFT normalization used in Sections~\ref{sec:CFT} and~\ref{sec:integrals} above. We then turn on a linearized backreaction of the scalar field on the geometry in order to compute the leading-order perturbation of the geometry.  The result allows us to compute~$v_\text{B}$, which we compare to the skipped pole results found above.

In order to more directly connect to the results in the previous sections, in this section we will work in terms of the conformal weight~$h$ of the perturbing operator~$\Ocal(x)$, rather than the conformal dimension~$\Delta$ that typically appears in the holographic literature, using the relationship~$\Delta = h + \bar{h} = 2h$ (since we are considering spin-zero fields,~$h = \bar{h}$).

\subsection{Massless Scalar on BTZ}
\label{eq:BTZscalar}

The planar BTZ black hole is (working in AdS units~$\ell = 1$)
\be
\label{eq:BTZ}
ds^2 = -(r^2-r_0^2) dt^2 + \frac{dr^2}{r^2 - r_0^2} + r^2 dx^2,
\ee
and its temperature is~$T = r_0/2\pi$.  The equation of motion for a massive scalar field is
\be
(-\grad^2 + m^2) \phi = 0,
\ee
where the mass~$m$ is related to the weight~$h$ as~$m^2 = 4h(h-1)$.  The relevant deformations~$h \in (0,1)$ we have been considering correspond to tachyonic bulk fields above (or equal to) the Breitenlohner-Freedman (BF) bound:~$m^2 \in [-1,0)$.  The static and planar-symmetric solution to the scalar equation of motion that is regular on the BTZ horizon is
\begin{multline}
\label{eq:phihgeneral}
\phi = C \left[\frac{\Gamma(1-h)^2}{\Gamma(2(1-h))} \rho^{-2(1-h)} \pFq{2}{1}{1-h,1-h}{2(1-h)}{\frac{1}{\rho^2}} \right. \\ \left. -\frac{\Gamma(h)^2}{\Gamma(2h)} \rho^{-2h} \pFq{2}{1}{h,h}{2h}{\frac{1}{\rho^2}} \right],
\end{multline}
where~$\rho \equiv r/r_0$ and~$C$ is an arbitrary constant.

In order to eventually match with the CFT results, we must express~$C$ in terms of the perturbation parameter~$\lambda$ introduced in Section~\ref{sec:CFT}.  This can be accomplished by enforcing that the boundary operator dual to the bulk field~$\phi$ have the same normalization as the operator~$\mathcal{O}(x)$ used in the CFT analysis. In the CFT analysis, the action was deformed as
\be
\label{eq:CFTdeform}
S \to S + \int d^2 u \, \lambda(u) \mathcal{O}(u)
\ee
where the source~$\lambda(u)$ was constant and~$\mathcal{O}(u)$ was normalized to have the vacuum two-point function
\be
\label{eq:CFT2pt}
\ev{\mathcal{O}(u)\mathcal{O}(v)}_0 = \frac{1}{|u-v|^{4h}}.
\ee
On the other hand, in holography turning on a bulk scalar field corresponds to deforming the boundary action as
\be
\label{eq:holodeform}
S \to S + \int d^2 u \, \phi_{(0)}(u) \mathcal{O}_\phi(u),
\ee
where the source~$\phi_0(u)$ and the expectation value of the operator~$\mathcal{O}_\phi(u)$ correspond to the near-boundary behavior of the bulk scalar~$\phi$: in Fefferman-Graham coordinates~$(z,u)$ (with~$u$ and~$\bar{u}$ complex coordinates~$u = x +i\tau = x - t$,~$\bar{u} = x - i\tau = x + t$ as usual), we have~\cite{Klebanov:1999tb,Skenderis:2002wp}
\be
\phi(z,u) = z^{2(1-h)} \left[\phi_{(0)}(u) + O(z^2)\right] + z^{2h} \left[\frac{1}{2(1-2h)} \ev{\mathcal{O}_\phi(u)}_s + O(z^2)\right],
\ee
where the subscript on~$\ev{\mathcal{O}_\phi(u)}_s$ denotes that the one-point function is computed in the theory with the source~$\phi_0(u)$ turned on.  The vacuum two-point function of the operator~$\mathcal{O}_\phi(u)$ is given by~\cite{Witten:1998qj,Freedman:1998tz}
\be
\label{eq:holo2pt}
\ev{\mathcal{O}_\phi(u)\mathcal{O}_\phi(v)}_0 = \frac{2(2h - 1)^2}{\pi} \frac{1}{|u-v|^{4h}}.
\ee
Comparing~\eqref{eq:CFT2pt} and~\eqref{eq:holo2pt} indicates that the holographic and CFT operators are related by
\be
\mathcal{O}_\phi(u) = \sqrt{\frac{2}{\pi}} \, (1-2h) \mathcal{O}(u),
\ee
and therefore in order for the deformations~\eqref{eq:CFTdeform} and~\eqref{eq:holodeform} to match, the holographic and CFT source functions must be related via
\be
\label{eq:phi0lambda}
\phi_{(0)}(u) = \sqrt{\frac{\pi}{2}} \, \frac{1}{1-2h} \, \lambda(u).
\ee

A near-boundary expansion of~\eqref{eq:phihgeneral} gives\footnote{The BTZ metric~\eqref{eq:BTZ} is put into Fefferman-Graham form using~$r = (1+(r_0 z/2)^2)/z$.}
\be
\phi_{(0)} = C \frac{\Gamma(1-h)^2}{\Gamma(2(1-h))} r_0^{2(1-h)},
\ee
which when combined with~\eqref{eq:phi0lambda} gives us the matching
\be
C = r_0^{2(h-1)} \sqrt{\frac{\pi}{2}} \frac{\Gamma(2(1-h))}{(1-2h)\Gamma(1-h)^2} \, \lambda.
\ee
So in terms of CFT parameters~$\lambda$ and~$\beta$, the bulk scalar solution is
\begin{multline}
\label{eq:philambda}
\phi = \frac{\sqrt{\pi/2} \, \lambda}{1-2h} \left(\frac{2\pi}{\beta}\right)^{2(h-1)} \left[\rho^{-2(1-h)} \pFq{2}{1}{1-h,1-h}{2(1-h)}{\frac{1}{\rho^2}} \right. \\ \left. -\frac{\Gamma(h)^2 \Gamma(2(1-h))}{\Gamma(1-h)^2\Gamma(2h)} \rho^{-2h} \pFq{2}{1}{h,h}{2h}{\frac{1}{\rho^2}} \right].
\end{multline}

\subsection{Backreaction on the Geometry}
\label{subsec:backreaction}

Having found the bulk scalar solution on the BTZ background, we next allow it to backreact on the geometry in order to compute the perturbed butterfly velocity.  The full Einstein-scalar-AdS equations are
\be
R_{ab} - \frac{1}{2} R g_{ab} - g_{ab} = 8\pi G_\text{N} T_{ab},
\ee
where the scalar field stress tensor is
\be
T_{ab} = \grad_a \phi \grad_b \phi - \frac{1}{2} g_{ab} \left(\grad^c \phi \grad_c \phi + m^2 \phi^2\right),
\ee
where as above~$m^2 = 4h(h-1)$.  We then look for solutions of the form
\be
ds^2 = -r_0^2\left[\rho^2-1 + p(\rho)\right] dt^2 + \frac{d\rho^2}{\rho^2-1 + p(\rho)} + r_0^2 \rho^2 \left[1 + q(\rho)\right] dx^2.
\ee
Treating~$p(\rho)$ and~$q(\rho)$ as perturbations of order~$\phi^2$ (and hence of order~$\lambda^2$), the linearized Einstein-scalar-AdS equations become
\begin{subequations}
\begin{align}
\left(\rho^2 q'(\rho)\right)' &= -16\pi G_\text{N} \rho^2 \phi'(\rho)^2, \label{subeq:qeq} \\
p'(\rho) &= - \rho^2 q'(\rho) + 8\pi G_\text{N} \rho \left[4h(1-h)\phi(\rho)^2 +(\rho^2-1)\phi'(\rho)^2\right], \label{subeq:peq}
\end{align}
\end{subequations}
where primes denote derivatives with respect to~$\rho$.

The equations of motion for~$p(\rho)$ and~$q(\rho)$ can be straightforwardly integrated once appropriate boundary conditions are chosen.  It is convenient to make the gauge choice that the horizon remains at~$\rho = 1$ even after the perturbation, which imposes~$p(1) = 0$.  It is also convenient to enforce that in the perturbed system, the relationship between~$r_0$ and the black hole temperature be the same as in the background BTZ solution:~$T = r_0/2\pi$.  This imposes~$p'(1) = 0$.  Finally, in order to leave the boundary metric unchanged we must demand that~$q(\rho) \to 0$ as~$\rho \to \infty$ (and likewise that~$p(\rho)$ grow more slowly than~$\rho^2$).  We therefore take as our boundary conditions
\be
p(1) = 0 = p'(1), \quad q(\infty) = 0.
\ee
The equation of motion~\eqref{subeq:peq} for~$p(\rho)$ implies that if~$p'(1) = 0$, then
\be
q'(1) = 32\pi G_\text{N} h(1-h) \phi(1)^2.
\ee
We can then integrate~\eqref{subeq:qeq} imposing these boundary conditions to obtain
\be
\label{eq:qgrav}
q(\rho) = \frac{16\pi G_\text{N}}{\rho} \left[\int_1^\infty d\rho_1 \, \rho_1 \min(\rho,\rho_1) \phi'(\rho_1)^2 - 2h(1-h) \phi(1)^2\right].
\ee
As we will see, this is all that's needed to find the butterfly velocity.

\subsection{The Butterfly Velocity}
\label{subsec:butterfly}

The butterfly velocity for a general class of metrics of the form
\be
\label{eq:generalBH}
ds^2 = -f(r) dt^2 + \frac{dr^2}{f(r)} + s(r) dx^2
\ee
was computed in~\cite{Blake:2016wvh}: it is
\be
v_\text{B} = \sqrt{\frac{4\pi T}{s'(r_0)}} = \sqrt{\frac{f'(r_0)}{s'(r_0)}},
\ee
where~$r_0$ is the largest root of~$f(r)$ (and the black hole temperature is~$T = f'(r_0)/4\pi$, as usual).  The Lyapunov exponent can also be computed, but it is less interesting since it is always dominated by the gravitational contribution:~$\lambda_\mathrm{L} = 2\pi T = f'(r_0)/2$.

\begin{figure}[t]
\centering
\includegraphics[width=0.5\textwidth]{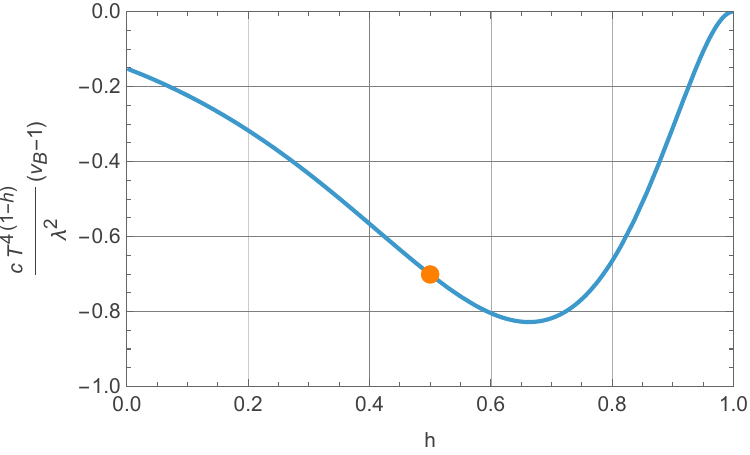}
\caption[]{The perturbation to the butterfly velocity as a function of~$h$.  The blue curve shows the result~\eqref{eq:vPSWard} for the pole-skipping velocity obtained from the Ward identities, as well as the holographic result~\eqref{eq:vBphi} for the butterfly velocity obtained from the holographic calculation (the results coincide).  The orange point shows the CFT result obtained in Section~\ref{sec:integrals} for the special case~$h = 1/2$.  Near~$h = 1$, the CFT results predict that~$(v_\text{B}-1)$ should be quadratic in~$(h-1)$, also consistent with the holographic results shown here.}
\label{fig:vB}
\end{figure}

In our case,~$f(r) = r^2 - r_0^2 + r_0^2 p(r/r_0)$ and~$s(r) = r^2 (1+q(r/r_0))$.  Using the solution~\eqref{eq:qgrav} and the fact that~$p'(1) = 0$, we obtain a butterfly velocity of
\be
\label{eq:vBphi}
v_\text{B} = \frac{1}{\sqrt{1+q(1)+ q'(1)/2}} = 1 - \frac{12\pi}{c} \left[\int_1^\infty d\rho \, \rho \, \phi'(\rho)^2 - h(1-h) \phi(1)^2\right] + O(\lambda^3),
\ee
where we used~$G_\text{N} = 3/2c$ to express the result in terms of CFT quantities.  This expression can be straightforwardly integrated numerically using the solution~\eqref{eq:philambda} for the bulk scalar; the result is shown in Figure~\ref{fig:vB} for~$h \in (0,1)$, which agrees exactly with the result~\eqref{eq:vPSWard} for the pole-skipping veocity obtained from the Ward identities in Section~\ref{sec:Ward}.

Note that in~\eqref{eq:vPShone}, we found that the pole-skipping velocity near~$h = 1$ vanishes to~$O(\lambda^2(h-1)^2,\lambda^3)$, which is consistent with the behavior of the butterfly velocity shown in Figure~\ref{fig:vB}.  Likewise, the pole-skipping velocity for~$h = 1/2$ was found in~\eqref{eq:vPShhalf}, which is shown in the figure as an orange dot.  Note that it, too, agrees precisely with the holographic results\footnote{We also note that our result agrees with the holographic calculation presented in \cite{Jiang:2024tdj} for $h=3/4$.}.

\section{Conclusion}
\label{sec:conclusion}

We have described how the location of skipped poles in the stress tensor two-point function of a general two-dimensional CFT is modified upon the deformation of the theory by a relevant operator of conformal weight~$h$. We first obtained the leading order deformation of the skipped poles and the corresponding velocity $v_{\mathrm{PS}}$ for generic $h$ by use of the diffeomorphism and Weyl Ward identities.  We then obtained the same results by a direct calculation in conformal perturbation theory. In this direct approach, we were able to find the leading but we only analyzed the Fourier transform and skipped poles for the special cases~$h = 1/2$ and~$h$ near~$1$. To verify our results, we utilized the fact that the leading-order perturbation to the two-point function is universal in the CFT central charge, and hence it must match the result obtained from a holographic dual.  Indeed, a holographic calculation of the butterfly velocity in the deformed theory matches the pole-skipping velocity we obtained from conformal perturbation theory.  This matching is expected on general grounds in a holographic theory, but it provides a highly nontrivial verification that our distributional approach to the conformal perturbation theory is correct.

Our results therefore provide a first step towards exploring pole skipping phenomenon away from holographic limits and conformality.  By combining the two approaches used in this work, one can hope to obtain the higher order corrections to $v_{\mathrm{PS}}$: the discussion in Section~\ref{sec:Ward} illustrates that the order~$\lambda^3$ correction to the two-point function will be determined by the order $\lambda$ correction to the scalar two-point function, $G_{1, \Ocal \Ocal}$, as well as the order $\lambda^2$ correction to $\langle \Ocal \rangle$. Using the techniques presented in this work, we expect the calculation of these quantities to be relatively tractable. It will be much more difficult to apply conformal perturbation theory to compute OTOCs in the deformed CFT in order to obtain the deformed Lyapunov exponent and butterfly velocity of the field theory (note that these need not coincide with the holographic results).  Such OTOCs are more difficult to compute: an OTOC is constructed from at least a four point function, but computing it to order~$\lambda^2$ in the deformed CFT would require two more insertions of the perturbing operator, leading to a correlation function of at least six points.  We leave such a calculation to future work, though we note recent results on higher point functions in CFT and conformal perturbation theory, such as \cite{Burrington:2023vei}, may be helpful in such calculations.

Beyond validating the holographic AdS/CFT correspondence away from strict conformality, conformal perturbation theory has physical significance in terms of comparison to lattice studies of a trapped Ising models, SU($N$) Yang-Mills theory and even of QCD in certain regimes \cite{Caselle:2019tiv, Amoretti:2020bad, Caselle:2024zoh}. Our study contributes to a more quantitative understanding of quantum chaos in such perturbed systems, which may have relevance for recent studies of chaos in quantum spin systems \cite{Craps:2019rbj, Nizami:2024ltk} and for experimental systems that realize quantum spin chains near a critical point, e.g., \cite{Jepsen:2020cjl, Periwal:2021eur, Zhao:2024dsh, Bamba:2023ohp, Reinic:2024oox, Jeffrey:2026oyb}.

Other interesting generalizations of the work here would be to consider more general deformation operators, such as those with non-zero spin ($h \neq \bar{h}$), to work in higher dimensions, to consider perturbed CFTs on the torus and higher genus Riemann surfaces, and to track conformal deformations beyond perturbation theory, say along an entire renormalization group trajectory between two fixed points, as in, e.g., \cite{Cappelli:1989yu, Karateev:2024skm}.  We leave all this to future work.

\section*{Acknowledgments}

We thank Paolo Arnaudo, Richard Davison, Edgar Shaghoulian, Shreya Vardan, and Benjamin Withers for useful discussions. This research is supported in part by the U.S. Department of Energy, Office of Science, Office of High Energy Physics RENEW-HEP program, under Award Numbers DE-SC 0023876 (A. Miller) and DE-SC 0024518 (C. Asplund).  SF is supported by an APS-Simons Travel and Professional Development Award. DMR is partially supported by STFC consolidated grant ST/X000664/1 and by the Simons Investigator award \#620869.

\appendix

\section{Thermodynamics in Conformal Perturbation Theory}
\label{app:cpt-thermo}

In this Appendix, we review a few aspects of the thermodynamics of a
CFT perturbed by a relevant deformation in order to obtain the unperturbed correlators needed for the analysis in Section~\ref{sec:Ward}. As discussed in Section~\ref{sec:CFT}, the partition function of the deformed theory on the cylinder is given by:
\begin{align}
  {Z[\lambda] \over Z^{(0)}}={}& \left\langle \exp\left[{-} \lambda \int_{\text{cyl}} \dd^2 \tilde{z} \, {\cal O}(\tilde{z}) \right]\right\rangle_0 \nonumber \\
  ={}& 1 - \lambda \int \dd^2 \tilde{z} \, \left\langle {\cal O}(\tilde{z}) \right\rangle_0 + {\lambda^2 \over 2} \int \dd^2 \tilde{z}_1 \dd^2 \tilde{z}_2 \, \left\langle {\cal O}(\tilde{z}_1) {\cal O}(\tilde{z}_2) \right\rangle_0 + \dotsb\, .
\end{align}
As we are working in the canonical ensemble, this leads to the free
energy (density):
\begin{align}
  \beta f[\lambda] ={}& \beta f_0 + \beta \lambda \langle {\cal O}\rangle_0 - {\beta \lambda^2 \over 2} \int \dd^2 \tilde{z}\, \left\langle {\cal O}(\tilde{z}) {\cal O}(0)\right\rangle^{(c)}_0 + \dotsb\, .
\end{align}
Here we have used translation invariance to perform one of the
integrals for each term, and $f_0 = - {\pi c \over 6\beta^2}$ is the
unperturbed free energy density. Assuming that ${\cal O}$ is a
Virasoro primary, the one-point function $\langle{\cal O}\rangle_0$
vanishes on the cylinder and we are left with a single integral to
perform:
\begin{align}
  \int \dd^2 \tilde{z} \, \left\langle {\cal O}(\tilde{z}) {\cal O}(0)\right\rangle_0 = G_{0,{\cal O}{\cal O}}^\text{E}(\omega_n=0,k=0)\, ,
\end{align}
where for the second equality we've recognized the Fourier transform
of the Euclidean Green's function
$G_{0,{\cal O}{\cal O}}^\text{E}(\tilde{x},\tilde{\tau})$. From the free energy
density, we can readily obtain the entropy density $s$ via
$s = - \partial_T f$ and hence the enthalpy density
$\varepsilon +p = s T = - T \partial_T f$, where $\varepsilon$ is the
energy density and $p = - f$ is the pressure.

In fact, this same integral governs the leading order correction to
the one-point function of ${\cal O}$. The perturbative expansion for
$\langle {\cal O}\rangle$ reads
\begin{align}
  \langle {\cal O}(0) \rangle ={}& {\left\langle {\cal O}(0) \exp \left[ - \lambda \int \dd^2 \tilde{z} {\cal O}(\tilde{z}) \right]\right\rangle_0 \over \left\langle \exp \left[ - \lambda \int \dd^2 \tilde{z} {\cal O}(\tilde{z}) \right]\right\rangle_0} = \langle {\cal O}(0) \rangle_0 - \lambda \int \dd^2 \tilde{z} \left\langle {\cal O}(\tilde{z}) {\cal O}(0) \right\rangle^{(c)}_0 + \dotsb\, ,
\end{align}
or, upon using $\langle {\cal O}\rangle_0 = 0$,
\begin{align}
  {\langle {\cal O}\rangle \over \lambda} ={}& {-} \int \dd^2 \tilde{z} \left\langle {\cal O}(\tilde{z}) {\cal O}(0) \right\rangle_0 = \left. \partial_\lambda^2 f\right\vert_{\lambda=0} + \dotsb\, .
  \label{eq:Ovev-order1}
\end{align}

As is well-known and discussed in the main text, the Euclidean Green's
function for a Virasoro primary on the cylinder in the unperturbed CFT can be obtained via a
conformal transformation from the plane to the cylinder:
\begin{align}
  G_{0,{\cal O}{\cal O}}^\text{E}(\tilde{x}, \tilde{\tau}) ={}& \left(\frac{\pi}{\beta}\right)^{4h} \frac{1}{\left[\sinh \left({\pi \tilde{z} \over \beta} \right) \sinh \left({\pi \bar{\tilde{z}} \over \beta} \right) \right]^{4h}}\, , 
\end{align}
where $\tilde{z} = \tilde{x} + i \tilde{\tau}$ and
$\bar{\tilde{z}} = \tilde{x} - i \tilde{\tau}$. The Fourier
transform of this expression can be computed \cite{SSS-PRB} and
reads:
\begin{align}
\label{eq:GEOO-FT}
  G_{0,{\cal O}{\cal O}}^\text{E}(\omega_n, k) ={}& \pi \left(\frac{2\pi}{\beta}\right)^{4h-2} \frac{\Gamma(1-2h)}{\Gamma(2h)} \left\vert\frac{\Gamma \left(h+ \frac{\beta}{4\pi}(\abs{\omega_n} + i k) \right)}{\Gamma \left(1 - h + \frac{\beta}{4\pi}(\abs{\omega_n} + i k)\right)} \right\vert^2\, .
\end{align}
Here $\omega_n = {2\pi n \over \beta}$ are the Matsubara frequencies appropriate for a (bosonic) correlator on the thermal circle of radius $\beta$. We note that the integrals leading to this result converge directly
for $0< 2h <1$, and we are analytically continuing the result for general $h$.

With this integral, the enthalpy density to order $\lambda^2$ is
therefore
\begin{align}
\label{eq:enthalpy}
  \varepsilon + p ={}& s T = {-} T \partial_T f = \frac{\pi c}{3\beta^2} - \pi \lambda^2 \left(\frac{2\pi}{\beta} \right)^{4h-2} {\Gamma(2-2h) \over \Gamma(2h)} {\Gamma \left(h \right)^2 \over \Gamma\left(1 - h\right)^2} + \dotsb \, ,
\end{align}
while $\langle {\cal O}\rangle$ is given by
\begin{align}
\label{eq:Oonepoint}
  \langle {\cal O} \rangle ={}& {-}\lambda \pi \left( \frac{2\pi}{\beta} \right)^{4h-2} \frac{\Gamma(1-2h)}{\Gamma(2h)} \frac{\Gamma \left(h \right)^2}{\Gamma\left(1 - h \right)^2} + \dotsb\, .
\end{align}

Finally, the last ingredient for obtaining the leading order
correction to~$G_{0,--,--}$ in the approach of Section~\ref{sec:Ward} is
the retarded Green's function $G_{0,{\cal O}{\cal O}}(\omega,k)$ in
the unperturbed CFT. Fortunately, $G_{0,{\cal O}{\cal O}}(\omega,k)$
is obtained, up to a sign, by analytically continuing $G_{0,{\cal O}{\cal O}}^E(\omega_n,k)$ for $n>0$\footnote{In the case at hand, this amounts to taking \eqref{eq:GEOO-FT} for $n>0$, replacing $i\omega_n \to \omega + i \delta$ where $\delta$ is a positive infinitesimal, and multiplying by an overall negative sign (due to our conventions for $G^R$ and $G^E$).}, leading to:
\begin{align}
\label{eq:GROO-CFT}
  G_{0,{\cal O}{\cal O}}(\omega,k) ={}& {-}\pi \left(\frac{2\pi}{\beta} \right)^{4h-2} \frac{\Gamma(1-2h)}{\Gamma(2h)} \frac{\Gamma \left(h - i \frac{\beta}{4\pi}(\omega - k)\right) \Gamma \left(h - i \frac{\beta}{4\pi}(\omega + k) \right)}{\Gamma \left(1 - h - i \frac{\beta}{4\pi}(\omega - k)\right) \Gamma \left(1- h - i \frac{\beta}{4\pi}(\omega + k)\right)}\, .
\end{align}

Finally, we use the results of this appendix to compute the quadratic correction to $G_{--,--}$ using \eqref{eq:GR-full}: 
\begin{align}
G_{2,--,--} ={}& \frac{(1-h)^2}{4} \left(\frac{\omega+k}{\omega-k} \right)^2 G_{0,\Ocal\Ocal}(\omega,k) \label{eq:GR2-WI}\\
&- \frac{\pi}{8} (\omega+k)\left( \frac{\pi}{\beta} \right)^{2(2h-1)} \frac{\omega(h^2-3h+1) + k(h^2+h-1)}{(\omega-k)^2}\frac{\Gamma(\frac{1}{2}-h)\Gamma(h)}{\Gamma(\frac{1}{2}+h)\Gamma(1-h)}\, . \nonumber
\end{align}

\section{Distributions}
\label{app:distributions}

In this Appendix, we provide more details on the distributional interpretations of the integrals appearing in Section~\ref{sec:integrals}.

\subsection{Principal Value Form of $\Dcal^\text{H}(1/z^2)$}
\label{subapp:principalvalue}

To prove the equality~\eqref{eq:DHprincipalvalue}, we start by noting that since
\be
\label{eq:zn2ibp}
\int_{\mathbb{C}} d^2 z \, \varphi(z,\zb) \Dcal^\text{H}\frac{1}{z^2} \equiv \int d^2 z \, \frac{\partial_z \varphi(z,\zb)}{z}
\ee
is manifestly convergent, we can write
\begin{subequations}
\begin{align}
\int d^2 z \, \frac{\partial_z \varphi(z,\zb)}{z} &= \lim_{\eps \to 0} \int_{\mathbb{C} \setminus B_\eps(0)} d^2 z \, \frac{\partial_z \varphi(z,\zb)}{z} \\
    &= \lim_{\eps \to 0} \int_{\mathbb{C} \setminus B_\eps(0)} d^2 z \left[ \frac{\varphi(z,\zb)}{z^2} + \partial_z \left(\frac{\varphi(z,\zb)}{z}\right)\right],
\end{align}
\end{subequations}
where the product rule to get the last expression is allowed because the region of integration stays away from the pole.  The second term can be integrated using Stokes' theorem:
\be
\lim_{\eps \to 0} \int_{\mathbb{C} \setminus B_\eps(0)} d^2 z \, \partial_z \left(\frac{\varphi(z,\zb)}{z}\right) = \lim_{\eps \to 0} \frac{1}{2i} \int_{C_\eps(0)} d \zb \, \frac{\varphi(z,\zb)}{z},
\ee
where~$C_\eps(0)$ is a circle of radius~$\eps$ centered on the origin.  Setting~$z = \eps e^{i\theta}$ and~$\zb = \eps e^{-i\theta}$ to perform the integral, we find that in the limit~$\eps \to 0$ the integral vanishes, and hence~\eqref{eq:zn2ibp} reduces to~\eqref{eq:DHprincipalvalue} as claimed.

\subsection{Interpretations of $\sgn(x)^{n-1}/|x|^n$}
\label{subapp:absvalue}

To start, consider the function~$1/|x|$, which is defined on~$\mathbb{R} \setminus \{0\}$.  A standard way of extending this function to a distribution on the entire real line is by noting that since away from the origin~$1/|x| = (\sgn(x)\ln|x|)'$, we can define the distribution~$\Dcal(1/|x|)$ as a distributional derivative:
\be
\Dcal\frac{1}{|x|} = (\sgn(x) \ln|x|)',
\ee
meaning that the action of the distribution~$\Dcal(1/|x|)$ on a test function~$\varphi(x)$ is given by~\cite{GelShi}
\be
\int_{-\infty}^\infty \varphi(x) \Dcal\frac{1}{|x|} \, dx \equiv -\int_{-\infty}^\infty \varphi'(x) \sgn(x) \ln |x| \, dx.
\ee
This can be put into a slightly more convenient form by splitting up the integral into regions~$|x| < \eps$ and~$|x| > \eps$ for an arbitrary~$\eps > 0$, then writing~$\varphi'(x) = (\varphi(x)-\varphi(0))'$ and integrating by parts:
\begin{subequations}
\begin{align}
\int_{-\infty}^\infty \varphi(x) \Dcal\frac{1}{|x|} \, dx &= \int_{|x| \geq \eps} \frac{\varphi(x)}{|x|} \, dx + \int_{|x| < \eps} \frac{\varphi(x)-\varphi(0)}{|x|} \, dx + 2 \varphi(0) \ln \eps, \\
		&= \lim_{\eps \to 0} \left[\int_{|x| \geq \eps} \frac{\varphi(x)}{|x|} \, dx + 2 \varphi(0) \ln \eps \right], \label{subeq:oneoverabsx}
\end{align}
\end{subequations}
where the second line is obtained by taking~$\eps \to 0$.  Note that the second line is essentially a na\"ive regulated integral: we cut off the integral at~$x = \pm \eps$ and subtract off the divergent pieces as~$\eps \to 0$.  For example, the Fourier transform of~$\Dcal(1/|x|)$ is
\be
\label{eq:absxFourier}
\int_{-\infty}^\infty e^{-ik x} \Dcal \frac{1}{|x|} \, dx = \lim_{\eps \to 0} \left[2 \int_\eps^\infty \frac{\cos(kx)}{x} \, dx + 2\ln \eps\right] = -2\left(\ln |k| +\gamma\right),
\ee
where~$\gamma$ is the Euler-Mascheroni constant.

For functions of the form~$\sgn(x)^{n-1}/|x|^n$ with~$n$ a positive integer, a natural generalization of the procedure above is to define
\be
\label{eq:absndist}
\Dcal \frac{\sgn(x)^{n-1}}{|x|^n} = \frac{(-1)^{n-1}}{(n-1)!} \frac{d^n}{dx^n}\left(\sgn(x)\ln|x|\right).
\ee
For example, we will specifically need the~$n = 2$ and~$n = 3$ cases, for which
\begin{subequations}
\begin{align}
\int_{-\infty}^\infty \varphi(x) \Dcal \frac{1}{x|x|} \, dx &\equiv -\int_{-\infty}^\infty \varphi''(x) \sgn(x) \ln |x| \, dx \\
		&= \int_{|x| \leq \eps} \frac{\varphi(x) - (\varphi(0) + x \varphi'(0))}{x|x|} + \int_{|x| > \eps} \frac{\varphi(x)}{x|x|} + 2\varphi'(0) (\ln \eps - 1), \label{subeq:Dxsquaredfiniteeps} \\
		&= \lim_{\eps \to 0}\left[\int_{|x| > \eps} \frac{\varphi(x)}{x|x|} + 2\varphi'(0) (\ln \eps - 1)\right], \label{subeq:Dxsquared} \\
\int_{-\infty}^\infty \varphi(x) \Dcal \frac{1}{|x|^3} \, dx &\equiv -\frac{1}{2} \int_{-\infty}^\infty \varphi'''(x) \sgn(x) \ln |x| \, dx \\
		&= \int_{|x| \leq \eps} \frac{\varphi(x) - (\varphi(0) + \varphi'(0) x + \varphi''(0) x^2/2)}{|x|^3} + \int_{|x| > \eps} \frac{\varphi(x)}{|x|^3} \, dx \nonumber \\
		&\hspace{5cm} - \frac{\varphi(0)}{\eps^2} + \varphi''(0)\left(\ln \eps - \frac{3}{2}\right), \label{subeq:Dxcubedfiniteeps} \\
		&= \lim_{\eps \to 0}\left[\int_{|x| > \eps} \frac{\varphi(x)}{|x|^3} \, dx - \frac{\varphi(0)}{\eps^2} + \varphi''(0)\left(\ln \eps - \frac{3}{2}\right)\right], \label{subeq:Dxcubed}
\end{align}
\end{subequations}
where in~\eqref{subeq:Dxsquaredfiniteeps} and~\eqref{subeq:Dxcubedfiniteeps},~$\eps > 0$ is arbitrary.  These distributions have Fourier transforms
\be
\int_{-\infty}^\infty e^{-ikx} \Dcal\frac{1}{x|x|} \, dx = 2ik \left(\ln|k| + \gamma\right), \qquad \int_{-\infty}^\infty e^{-ikx} \Dcal\frac{1}{|x|^3} \, dx = k^2 \left(\ln|k| + \gamma\right).
\ee
In fact, since from~\eqref{eq:absndist} we have that for any positive integer~$n$
\be
\frac{d}{dx} \Dcal \frac{\sgn(x)^{n-1}}{|x|^n} = -n \Dcal \frac{\sgn(x)^n}{|x|^{n+1}},
\ee
the Fourier transforms of these distributions can be obtained inductively from that of~$\Dcal(1/|x|)$:
\be
\label{eq:DFourier}
\int_{-\infty}^\infty e^{-ikx} \Dcal\frac{\sgn(x)^{n-1}}{|x|^n} \, dx = -\frac{2(-ik)^{n-1}}{(n-1)!} \left(\ln|k| + \gamma\right).
\ee

Any other distributional extension of~$\sgn(x)^{n-1}/|x|^n$ to the entire real line must be related to~$\Dcal(\sgn(x)^{n-1}/|x|^n)$ by delta functions and derivatives thereof.  For instance, another natural extension is obtained through the following observations.  Note that away from~$x = 0$ we have~$x(\sgn(x)^{n-1}/|x|^n) = \sgn(x)^{n-2}/|x|^{n-1}$.  However, the Fourier transforms~\eqref{eq:DFourier} imply that this relationship does \textit{not} hold for the distributions~$\Dcal(\sgn(x)^{n-1}/|x|^n)$, since
\begin{multline}
\int_{-\infty}^\infty e^{-ikx} x \Dcal \frac{\sgn(x)^{n-1}}{|x|^n} \, dx = i \frac{d}{dk} \int_{-\infty}^\infty e^{-ikx} \Dcal \frac{\sgn(x)^{n-1}}{|x|^n} \, dx \neq \int_{-\infty}^\infty e^{-ikx} \Dcal \frac{\sgn(x)^{n-2}}{|x|^{n-1}} \, dx \\ \Rightarrow \quad x \Dcal \frac{\sgn(x)^{n-1}}{|x|^n} \neq \Dcal \frac{\sgn(x)^{n-2}}{|x|^{n-1}}.
\end{multline}
We can construct distributions that \textit{are} related by multiplication by~$x$ by successively integrating the Fourier transform of~$\Dcal(1/|x|)$ in~$k$.  Starting with~\eqref{eq:absxFourier}, induction leads to the distributions~$\widetilde{\Dcal}(\sgn(x)^{n-1}/|x|^n)$ whose Fourier transforms are
\be
\label{eq:DtildeFourier}
\int_{-\infty}^\infty e^{-ikx} \widetilde{\Dcal}\frac{\sgn(x)^{n-1}}{|x|^n} \, dx = -\frac{2(-ik)^{n-1}}{(n-1)!} \left(\ln|k| + \gamma - H_{n-1}\right),
\ee
where~$H_n$ is the~$n^\mathrm{th}$ harmonic number.  These \textit{do} have the property that
\be
i \frac{d}{dk} \int_{-\infty}^\infty e^{-ikx} \widetilde{\Dcal} \frac{\sgn(x)^{n-1}}{|x|^n} \, dx = \int_{-\infty}^\infty e^{-ikx} \widetilde{\Dcal} \frac{\sgn(x)^{n-2}}{|x|^{n-1}} \, dx,
\ee
and thus
\be
x \widetilde{\Dcal} \frac{\sgn(x)^{n-1}}{|x|^n} = \widetilde{\Dcal} \frac{\sgn(x)^{n-2}}{|x|^{n-1}}.
\ee
Note that because the difference between the Fourier transforms~\eqref{eq:DFourier} and~\eqref{eq:DtildeFourier} of $\Dcal(\sgn(x)^{n-1}/|x|^n)$ and~$\widetilde{\Dcal}(\sgn(x)^{n-1}/|x|^n)$ is an integer power of~$k$, these distributions are related by derivatives of the delta function, as expected:
\begin{subequations}
\begin{align}
\label{eq:DtildeD}
\widetilde{\Dcal} \frac{\sgn(x)^{n-1}}{|x|^n} &= \Dcal \frac{\sgn(x)^{n-1}}{|x|^n} + \frac{2(-1)^{n-1} H_{n-1}}{(n-1)!} \delta^{(n-1)}(x), \\
	&= \frac{(-1)^{n-1}}{(n-1)!} \frac{d^n}{dx^n}\left(\sgn(x)(\ln|x|+H_{n-1})\right).
\end{align}
\end{subequations}
(The second expression is obtained by using~\eqref{eq:absndist} and the fact that~$\sgn^{(n)}(x) = 2\delta^{(n-1)}(x)$.)  The specific distributions we need correspond to~$n = 2$ and~$n = 3$, whose action on an arbitrary test function can be obtained by combining~\eqref{eq:DtildeD} with e.g.~\eqref{subeq:Dxsquared} and~\eqref{subeq:Dxcubed}:
\begin{subequations}
\begin{align}
\int_{-\infty}^\infty \varphi(x) \widetilde{\Dcal} \frac{1}{x|x|} \, dx &= \lim_{\eps \to 0}\left[\int_{|x| > \eps} \frac{\varphi(x)}{x|x|} + 2\varphi'(0) \ln \eps\right], \\
\int_{-\infty}^\infty \varphi(x) \widetilde{\Dcal} \frac{1}{|x|^3} \, dx &= \lim_{\eps \to 0}\left[\int_{|x| > \eps} \frac{\varphi(x)}{|x|^3} \, dx - \frac{\varphi(0)}{\eps^2} + \varphi''(0)\ln \eps\right]. \label{subeq:Dtildex3}
\end{align}
\end{subequations}
In other words, to evaluate the action of these distributions on an arbitrary test function, we cut off the integral at~$x = \pm \eps$ and then just remove the divergent terms in~$\eps$.  This is completely analogous to the evaluation of the action of~$\Dcal(1/|x|)$.

\section{Integration Details}
\label{app:integrals}

In this Appendix, we provide more details on the evaluation of the integrals in Section~\ref{sec:integrals}.

\subsection{Stokes' Theorem}
\label{subapp:Stokes}

In evaluating the integrals we will make extensive use of the complex Stokes' theorem, which says that for any region~$\Omega$ of the complex plane and differentiable~$F(z,\zb)$ and~$G(z,\zb)$ on~$\Omega$, then\footnote{Recall that in our notation,~$d^2 z \equiv d\zb \wedge dz/2i$, hence the factor of~$1/2i$ on the right-hand side of~\eqref{eq:complexstokes}.}
\be
\label{eq:complexstokes}
\int_\Omega d^2 z \left(\partial_z F(z,\zb) + \partial_{\zb} G(z,\zb) \right) = \frac{1}{2i} \oint_{\partial\Omega} \left(F(z,\zb) d\zb + G(z,\zb) dz\right),
\ee
where the contour~$\partial \Omega$ should be traversed in the positive direction, meaning that as the variable of integration (either~$z$ or~$\zb$) runs along~$\partial \Omega$ then~$\Omega$ should be on the left-hand side.  See e.g.~\cite{Guida:1995kc} for an example of the use of Stokes' theorem in evaluating integrals over the complex plane.

\subsection{Evaluating $K(w,\Sigma)$}
\label{subapp:Kint}

\begin{figure}[t]
\centering
\includegraphics[page=3]{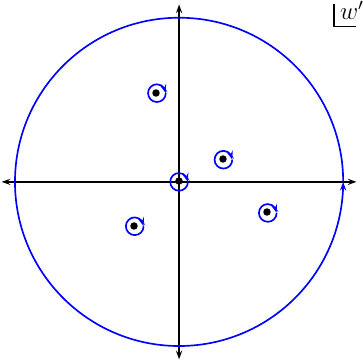}
\caption{The contour of integration~$\gamma$ for the integral~\eqref{eq:Kcontourint} obtained from applying Stokes' theorem.  The large circle is a contour at infinity traversed counterclockwise, while the five infinitesimal circles go clockwise around the logarithmic singularity at~$w' = 0$ and the four poles at~$w' = (\Sigma \pm 1)/(w \pm 1)$.}
\label{fig:Kcontour}
\end{figure}

To evaluate~$K(w,\Sigma)$, we first note that since away from the origin~$1/\bar{w}' = \partial_{\bar{w}'} \ln |w'|^2$, we can write~\eqref{eq:Kdef} as
\be
K(w,\Sigma) = \int_\Omega d^2 w' \, \partial_{\bar{w}'} \left\{\frac{w' \ln |w'|^2}{((\Sigma-w w')^2-(1-w')^2)^2} \left[1+\frac{8(h-1)w'^2}{((\Sigma-w w')^2-(1+w')^2)^2}\right]\right\},
\ee
where the region of integration~$\Omega$ excludes infinitesimal circles around the origin and the four poles at~$w' = (\Sigma \pm 1)/(w \pm 1)$.  Stokes' theorem then turns this integral into a contour integral:
\be
\label{eq:Kcontourint}
K(w,\Sigma) = \frac{1}{2i} \int_\gamma dw' \, \frac{w' \ln |w'|^2}{((\Sigma-w w')^2-(1-w')^2)^2} \left[1+\frac{8(h-1)w'^2}{((\Sigma-w w')^2-(1+w')^2)^2}\right],
\ee
where the contour~$\gamma$ contains infinitesimal circles around the five singular points as well as a contour around infinity, as shown in Figure~\ref{fig:Kcontour}.

Since the integrand of~\eqref{eq:Kcontourint} is not holomorphic in~$w'$, we cannot evaluate the contour integral using residues.  Instead, we evaluate the integrals directly.  For the circle at infinity, we write~$w' = R e^{i\theta}$, expand at large~$R$, integrate in~$\theta,$, and then take the limit~$R \to \infty$; this contribution vanishes.  Likewise, for each of the infinitesimal circles we write~$w' = w_* + \eps e^{i\theta}$ (where~$w_*$ is the center of the circle), expand at small~$\eps$, integrate in~$\theta$, and then take~$\eps \to 0$.  The contribution from the circle around the origin vanishes, while the contribution from the four poles gives the result~\eqref{eq:Kfinal} quoted in the main text.

\subsection{Evaluating $I(u,v)$}
\label{subapp:Iint}

To evaluate~$I(u,v)$, we again make use of Stokes' theorem.  First, note that by inserting the result~\eqref{eq:Kfinal} into the expression~\eqref{eq:IcKint}, we can rearrange some terms inside the integrand to reexpress~$I$ as
\be
\label{eq:IKtilde}
I = \frac{2^{7-4h}h}{(u-v)^4} \int_\mathbb{C} \frac{d^2 w}{|1-w^2|^{2(1-h)}} \, \widetilde{K}(w,\Sigma),
\ee
where now
\be
\widetilde{K}(w,\Sigma) \equiv \frac{\pi}{4} \left[-\frac{h-1}{(w-\Sigma)^2} + \frac{h(1-w\Sigma) + (h-1)(w-\Sigma)^2}{(w-\Sigma)^3} \ln\left|\frac{1}{\zeta} \, \frac{(w+1)}{(w-1)}\right|^2\right]
\ee
and where we have defined
\be
\zeta \equiv \frac{\Sigma+1}{\Sigma-1}.
\ee
Unlike the original~$K(w,\Sigma)$, which has poles at~$w = \pm \Sigma$,~$\widetilde{K}(w,\Sigma)$ only has a pole at~$w = \Sigma$, which simplifies the analysis.

\begin{figure}[t]
\centering
\includegraphics[page=4]{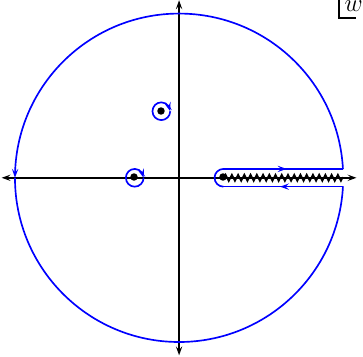}
\caption{The contour of integration~$\gamma$ for the integral~\eqref{eq:Icontour} obtained from applying Stokes' theorem.  The large circle is a contour at infinity traversed counterclockwise; it connects to a contour that runs around the branch cut along the half-axis~$w > 1$.  The two infinitesimal circles go around the singularities at~$w = -1$ and~$w = \Sigma$.}
\label{fig:Icontour}
\end{figure}

To use Stokes' theorem, note that it is possible to find an antiderivative of the integrand of~\eqref{eq:IKtilde} with respect to~$\bar{w}$:
\begin{multline}
I = \frac{2^{2(3-h)}}{(u-v)^4} \int_\Omega d^2 w \, \partial_{\bar{w}} \left\{\frac{1}{(1-w^2)^{1-h}} \left(\frac{1+\bar{w}}{1-\bar{w}}\right)^h \left[\widetilde{K}(w,\Sigma) \, \pFq{2}{1}{h,2h}{1+h}{\frac{\bar{w}+1}{\bar{w}-1}} \right. \right. \\ \left. \left. - \frac{\pi}{4h} \frac{h(1-w\Sigma) + (h-1)(w-\Sigma)^2}{(w-\Sigma)^3} \, \pFq{3}{2}{h,h,2h}{1+h,1+h}{\frac{\bar{w}+1}{\bar{w}-1}}\right]\right\}.
\end{multline}
The expression inside the curly braces is singular at~$w = \pm 1$ and~$w = X$, and also exhibits a branch cut along the half-axis~$w \in (1,\infty)$.  The region of integration~$\Omega$ must therefore exclude infinitesimal circles around~$w = -1$ and~$w = X$, as well as an infinitesimal neighborhood of~$(1,\infty)$.  With this in mind, applying Stokes' theorem then gives
\begin{multline}
\label{eq:Icontour}
I = \frac{2^{5-2h}}{(u-v)^4 i} \int_\gamma dw \, \frac{|1+w|^{2h}(1-w)^h}{(1-w^2)(1-\bar{w})^h} \left[\widetilde{K}(w,\Sigma)\, \pFq{2}{1}{h,2h}{1+h}{\frac{\bar{w}+1}{\bar{w}-1}} \right. \\ \left. - \frac{\pi}{4h} \frac{h(1-w\Sigma) + (h-1)(w-\Sigma)^2}{(w-\Sigma)^3} \, \pFq{3}{2}{h,h,2h}{1+h,1+h}{\frac{\bar{w}+1}{\bar{w}-1}}\right].
\end{multline}
where the contour~$\gamma$ contains infinitesimal circles around~$w = -1$ and~$w = X$ and a keyhole contour around~$w \in (1,\infty)$ that connects to a circle at infinity, as shown in Figure~\ref{fig:Icontour}.

The contributions from the various circles can be evaluated in the same way as we did for~$K(w,\Sigma)$.  The contributions from the circles at infinity and around~$w = \pm 1$ vanish, while the contribution from the pole at~$w = \Sigma$ gives a contribution of
\begin{multline}
I_\mathrm{pole} = \frac{2^{5-2h}\pi^2(h-1)}{(u-v)^4} \left(1-\Sigma^2\right)^{h-2} \left(-\bar{\zeta}\right)^h \left[(1+h) \Sigma \, \pFq{2}{1}{h,2h}{1+h}{\bar{\zeta}} \right. \\ \left. + \frac{(h-1)((1+2h)\Sigma^2 -1)}{2h} \, \pFq{3}{2}{h,h,2h}{1+h,1+h}{\bar{\zeta}}\right].
\end{multline}

To obtain the contribution from the portion of the contour that runs along the branch cut, we need to identify the discontinuity in the integrand of~\eqref{eq:Icontour} across the cut.  To do so, we make use of inversion formulas that relate~$_pF_q(z)$ to~$_pF_q(1/z)$ (see e.g.~\cite{NIST:DLMF}):
\begin{subequations}
\label{eqs:pFqinversions}
\begin{multline}
\pFq{2}{1}{a_1,a_2}{b_1}{z} = \frac{\Gamma(b_1)}{\Gamma(a_1)\Gamma(a_2)}\left[\frac{\Gamma(a_1)\Gamma(a_2-a_1)}{\Gamma(b_1-a_1)} (-z)^{-a_1} \pFq{2}{1}{a_1,a_1-b_1+1}{a_1-a_2+1}{\frac{1}{z}} \right. \\ \left. + \frac{\Gamma(a_2)\Gamma(a_1-a_2)}{\Gamma(b_1-a_2)} (-z)^{-a_2} \pFq{2}{1}{a_2,a_2-b_1+1}{a_2-a_1+1}{\frac{1}{z}}\right],
\end{multline}
\begin{multline}
\pFq{3}{2}{a_1,a_2,a_3}{b_1,b_2}{z} = \frac{\Gamma(b_1)\Gamma(b_2)}{\Gamma(a_1)\Gamma(a_2)\Gamma(a_3)} \\ \times \left[\frac{\Gamma(a_1)\Gamma(a_2-a_1)\Gamma(a_3-a_1)}{\Gamma(b_1-a_1)\Gamma(b_2-a_1)} (-z)^{-a_1}\pFq{3}{2}{a_1,a_1-b_1+1,a_1-b_2+1}{a_1-a_2+1,a_1-a_3+1}{\frac{1}{z}} \right. \\ 
\left. + \frac{\Gamma(a_2)\Gamma(a_1-a_2)\Gamma(a_3-a_2)}{\Gamma(b_1-a_2)\Gamma(b_2-a_2)} (-z)^{-a_2}\pFq{3}{2}{a_2,a_2-b_1+1,a_2-b_2+1}{a_2-a_1+1,a_2-a_3+1}{\frac{1}{z}} \right. \\
 \left. + \frac{\Gamma(a_3)\Gamma(a_1-a_3)\Gamma(a_2-a_3)}{\Gamma(b_1-a_3)\Gamma(b_2-a_3)} (-z)^{-a_3}\pFq{3}{2}{a_3,a_3-b_1+1,a_3-b_2+1}{a_3-a_1+1,a_3-a_2+1}{\frac{1}{z}} \right].
\end{multline}
\end{subequations}
The point is that if~$z$ is near the branch cut~$z \in (1,\infty)$ of~$_pF_q(z)$, then~$1/z$ stays away from the branch cut of~$_pF_q(1/z)$, and hence the discontinuity of~$_p F_q(z)$ across its branch cut can be computed by just finding the discontinuities across the branch cuts of the radicals~$(-z)^{a_i}$ in~\eqref{eqs:pFqinversions}.  In this way, we find that for real~$w > 1$
\begin{subequations}
\label{eqs:pFqdiscontinuities}
\bea
\pFq{2}{1}{h,2h}{1+h}{w \pm i\eps} &= -\frac{e^{\pm 2\pi i h}}{w^{2h}} \pFq{2}{1}{h,2h}{1+h}{\frac{1}{w}} + \frac{\Gamma(h)\Gamma(1+h)}{\Gamma(2h)} \frac{e^{\pm \pi i h}}{w^h}, \\
\pFq{3}{2}{h,h,2h}{1+h,1+h}{w \pm i \eps} &= \frac{e^{\pm 2\pi i h}}{w^{2h}} \pFq{3}{2}{h,h,2h}{1+h,1+h}{\frac{1}{w}} + \frac{\Gamma(1+h)^2}{\Gamma(2h)} \frac{e^{\pm \pi i h}}{w^h} \left(\ln w \mp \pi i\right).
\eea
\end{subequations}
Using these expressions, we find that the contribution of the branch cut to the integral is
\begin{multline}
I_\mathrm{cut} = \frac{\pi^2 \Gamma(1+h)}{2^{2(h-2)} (u-v)^4 \Gamma(1-h)\Gamma(2h)} \int_1^\infty dw \, \frac{(w^2-1)^{h-1}}{(w-\Sigma)^3} \left[(1-h)(w-\Sigma) \phantom{\frac{1}{2}} \right. \\ \left. + \left(h(1-w\Sigma)+(h-1)(w-\Sigma)^2\right)\left(\ln \left(\frac{1}{|\zeta|^2} \, \frac{w+1}{w-1}\right) - \pi \cot(\pi h)\right)\right],
\end{multline}
where we note that the integrand exhibits no branch cuts along~$w \in (1,\infty)$.  The radicals are relatively straightforward to integrate, but the logarithm poses more of a challenge.  To integrate it, we convert the logarithm into radicals using the fact that~$\ln z = \partial_\lambda z^\lambda |_{\lambda = 0}$.  We first define
\begin{multline}
I_\mathrm{cut}(\lambda) \equiv \frac{\pi^2 \Gamma(1+h)}{2^{2(h-2)} (u-v)^4 \Gamma(1-h)\Gamma(2h)} \int_1^\infty dw \, \frac{(w^2-1)^{h-1}}{(w-\Sigma)^3} \Bigg[(1-h)(w-\Sigma) \phantom{\frac{1}{2}} \\ + \left(h(1-w\Sigma)+(h-1)(w-\Sigma)^2\right)\left(\left(\frac{w+1}{w-1}\right)^\lambda - \ln |\zeta|^2 - \pi \cot(\pi h)\right)\Bigg],
\end{multline}
which converges for~$\lambda < h < 1$.  We therefore evaluate~$I_\mathrm{cut}(\lambda)$ for~$\lambda < h$, then take a derivative at~$\lambda = 0$ to obtain~$I_\mathrm{cut} = I'_\mathrm{cut}(0)$.  Proceeding in this way, we obtain
\begin{multline}
I = I_\mathrm{pole} + I_\mathrm{cut}'(0) = \frac{\pi^{7/2}\csc(2\pi h) \Gamma(1+h) ((1+2h)\Sigma^2-1)}{2^{2h-3} (u-v)^4(2-h) \Gamma(1-h)^2 \Gamma(h+\frac{1}{2}) \Gamma(2(h-1)) (1-\Sigma)^2 \Sigma} \\
	\times \left\{\left[-\ln\left|\zeta\right|^2 + \frac{2(1+h)\Sigma}{(1-h)((1+2h)\Sigma^2-1)}\right]\left[\pFq{2}{1}{2,h}{3-h}{\zeta} + \frac{(2-h)(1-\Sigma)^2}{2((1+2h)\Sigma^2-1)}\right] \right. \\
	\left. + \frac{(1-h)(2-h) \Sigma}{(2h-1)(1+\Sigma)} \left[ \frac{d}{d\lambda} \!\! \left. \pFq{2}{1}{1,h+\lambda}{2h}{\frac{2}{1-\Sigma}}\right|_{\lambda = 0} + \frac{(1-2h)(1-\Sigma)((2h^2-1)\Sigma^2+1)}{(1-h)^2((1+2h)\Sigma^2-1)^2} \right. \right. \\
	\left.\left. - \frac{\Gamma(1-h)\Gamma(h+\frac{1}{2})}{\sqrt{\pi}} (1-\Sigma)^h(-1-\Sigma)^{h-1} \left(-\ln\zeta + \pi \cot(\pi h)\right)\right] \right\} \\
	+ \frac{\pi^2}{2^{2h}} \frac{|1+\Sigma|^{2h} (1-\Sigma)^h}{(1-\Sigma^2)^2 (1-\overline{\Sigma})^h} \left\{2(h^2-1)\Sigma \pFq{2}{1}{h,2h}{1+h}{\bar{\zeta}} \right. \\ 
	\left. + \frac{(1-h)^2((1+2h)\Sigma^2-1)}{h} \pFq{3}{2}{h,h,2h}{1+h,1+h}{\bar{\zeta}}\right\}.
\end{multline}
Note that this expression contains a derivative of a hypergeometric function~$_2F_1$ with respect to one of its parameters.  To make the result more manageable, we eliminate this derivative by making use of the identity~\cite{Prudnikov:1990:IS}
\be
\pFq{3}{2}{a_1,a_2,a_3}{a_1+1,a_2+1}{z} = \frac{1}{a_2-a_1}\left[a_2 \pFq{2}{1}{a_1,a_3}{a_1+1}{z} - a_1 \pFq{2}{1}{a_2,a_3}{a_2+1}{z}\right].
\ee
Taking the limit~$a_2 \to a_1$ turns the right-hand side into a derivative of~$_2F_1$ with respect to its parameters, allowing us to express such derivatives of~$_2F_1$ in terms of~$_3F_2$.  Specifically, it is possible to show that
\begin{multline}
\frac{d}{d\lambda} \!\! \left. \pFq{2}{1}{1,h+\lambda}{2h}{\frac{2}{1-\Sigma}}\right|_{\lambda = 0} = \frac{2^{1-2h} \pi \Gamma(2h)}{\Gamma(h)^2} (1-\Sigma)^{2h} \left[ \frac{\csc(\pi h)}{\Sigma+1} \zeta^h \left(\ln \zeta - \pi \cot(\pi h)\right) \right. \\ \left. + \frac{2\cot(\pi h) \Gamma(2-2h)}{\Gamma(2-h)^2 (\Sigma-1)} \pFq{3}{2}{1-h,1-h,2(1-h)}{2-h,2-h}{\zeta}\right].
\end{multline}
Using this result to replace the derivative of~$_2F_1$ with a~$_3F_2$ eventually leaves us with
\begin{multline}
\label{eq:Ichatfinal}
I = \frac{4\pi^2 \Gamma(1+h)^2 \Gamma(1-2h)}{(u-v)^4 h\Gamma(1-h)^2 \Gamma(2h-2)} \frac{1}{1-\Sigma^2} \\
	\times \left[\left(\left(1-(1+2h)\Sigma^2\right)\ln\left|\zeta\right|^2 + \frac{2(1+h)}{1-h}\Sigma \right) \frac{(1-\Sigma)^{2(h-1)}}{2^{2h-1}} \pFq{2}{1}{1-h,2(1-h)}{2-h}{\zeta} \right. \\
	\left. +  \frac{(1-\Sigma)^{2(h-1)}(1-(1+2h)\Sigma^2)}{2^{2h-1}(h-1)} \pFq{3}{2}{1-h,1-h,2(1-h)}{2-h,2-h}{\zeta} - h \Sigma \ln|\zeta|^2 - \frac{1}{h-1} \right] \\
	+ \frac{\pi^2}{2^{2h}} \frac{|1+\Sigma|^{2h} (1-\Sigma)^h}{(1-\Sigma^2)^2 (1-\overline{\Sigma})^h} \left[2(h^2-1)\Sigma \pFq{2}{1}{h,2h}{1+h}{\bar{\zeta}} \right. \\ 
	\left. - \frac{(1-h)^2(1-(1+2h)\Sigma^2)}{h} \pFq{3}{2}{h,h,2h}{1+h,1+h}{\bar{\zeta}}\right],
\end{multline}
which yields the final expression~\eqref{eq:G2Eexplicit} for the perturbation to the Euclidean Green's function quoted in the text.

\section{Fourier Transform}
\label{app:Fourier}

This Appendix provides additional details on the Fourier transforms of Section~\ref{subsec:Fourier}.

\subsection{Branch cut discontinuities}
\label{subapp:tdisc}

We first need to identify the discontinuity of~$G_2^\text{E}(\Sigma(t,x),\overline{\Sigma}(t,x))$ across the real~$t$ axis.  To do so, first note that the logarithms in~\eqref{eq:G2Eexplicit} are single-valued in~$t$, since
\be
\ln\left|\frac{\Sigma+1}{\Sigma-1}\right|^2 = \ln \left(\frac{(\Sigma+1)(\overline{\Sigma}+1)}{(\Sigma-1)(\overline{\Sigma}-1)}\right) = \frac{4\pi x}{\beta}.
\ee
So these introduce no branch cuts in~$t$.  Any branch cuts instead come from the hypergeometric functions and the radicals; to study them, it is convenient to isolate the four terms in~\eqref{eq:G2Eexplicit} that exhibit cuts:
\begin{subequations}
\begin{align}
T_1(t,x) &\equiv (1-\Sigma)^{2(h-1)} \pFq{2}{1}{1-h,2(1-h)}{2-h}{\zeta}, \\
T_2(t,x) &\equiv (1-\Sigma)^{2(h-1)} \pFq{3}{2}{1-h,1-h,2(1-h)}{2-h,2-h}{\zeta}, \\
T_3(t,x) &\equiv \frac{(1+\overline{\Sigma})^h(1+\Sigma)^h(1-\Sigma)^h}{(1-\overline{\Sigma})^h} \pFq{2}{1}{h,2h}{1+h}{\bar{\zeta}}, \\
T_4(t,x) &\equiv \frac{(1+\overline{\Sigma})^h(1+\Sigma)^h(1-\Sigma)^h}{(1-\overline{\Sigma})^h} \pFq{3}{2}{h,h,2h}{1+h,1+h}{\bar{\zeta}},
\end{align}
\end{subequations}
where in terms of Lorentzian coordinates,
\be
\Sigma = \coth(\pi(x-t)/\beta), \quad \zeta = e^{2\pi(x -t)/\beta}, \quad \overline{\Sigma} = \coth(\pi(x+t)/\beta), \quad \bar{\zeta} = e^{2\pi(x +t)/\beta}.
\ee
In terms of the~$T_i(t,x)$, the discontinuity of~$G_2^\text{E}$ across the real~$t$ axis is
\begin{multline}
\label{eq:DeltaG2T}
\Delta G_2(t,x) \equiv G_2^\text{E}(\Sigma(t-i\eps,x),\overline{\Sigma}(t-i\eps,x)) - G_2^\text{E}(\Sigma(t+i\eps,x),\overline{\Sigma}(t+i\eps,x)) \\
= \frac{1}{2^{2h+1}} \left(\frac{2\pi}{\beta}\right)^{4h} \Bigg\{-\frac{\pi^2 \Gamma(1+h)^2 \Gamma(1-2h)}{2h\Gamma(1-h)^2 \Gamma(2h-2)} \csch^2(\xhat-\that) \\
	\times \left[\left(4\xhat\left(1-(1+2h)\coth^2(\xhat-\that)\right) + \frac{2(1+h)}{1-h}\coth(\xhat-\that) \right) \Delta T_1 +  \frac{1-(1+2h)\coth^2(\xhat-\that)}{h-1} \Delta T_2\right] \\
	+ \pi^2 \left[2(h^2-1)\coth(\xhat-\that) \Delta T_3 - \frac{(1-h)^2(1-(1+2h)\coth^2(\xhat-\that))}{h} \Delta T_4\right]\Bigg\},
\end{multline}
where~$\Delta T_i(t,x) \equiv T_i(t-i\eps,x) - T_i(t+i\eps,x)$ and we have defined~$\xhat \equiv \pi x/\beta$,~$\that \equiv \pi t/\beta$.  The discontinuities in the~$T_i$ can be obtained using the inversion formulas~\eqref{eqs:pFqdiscontinuities}.  We eventually find that
\begin{subequations}
\bea
\Delta T_1(t,x) &= \begin{cases} -\frac{2i\sin(\pi h)\Gamma(1-h)\Gamma(2-h)}{\Gamma(2(1-h))} \left[\csch(\xhat-\that)\right]^{2(h-1)}, & t < x, \\
		0, & t > x, \end{cases} \\
\Delta T_2(t,x) &= \begin{cases} \frac{2i \Gamma(2-h)^2}{\Gamma(2(1-h))} \left[\pi \cos(\pi h) - 2\sin(\pi h)(\xhat-\that) \right] \left[\csch(\xhat-\that)\right]^{2(h-1)}, & t < x, \\
		0, & t > x, \end{cases} \\
\Delta T_3(t,x) &= \begin{cases} 2i\sin(2\pi h) e^{2 h(\xhat+\that)} \left[\csch^2(\xhat-\that)\right]^{h} \pFq{2}{1}{h,2h}{1+h}{e^{2(\xhat+\that)}}, & t < -|x|, \\
		\Theta(x) \frac{2i\sin(\pi h)\Gamma(h)\Gamma(1+h)}{\Gamma(2h)} \left[\csch^2(\xhat-\that)\right]^h, & -|x| < t < |x|, \\
		 2i \left[\sin(2\pi h) e^{-2 h(\xhat+\that)} \pFq{2}{1}{h,2h}{1+h}{e^{-2(\xhat+\that)}} \phantom{\frac{\Gamma(h)}{\Gamma(2h)}} \right. & \\ \hfill \left. - \frac{\sin(\pi h)\Gamma(h)\Gamma(1+h)}{\Gamma(2h)}\right] \left[\csch^2(\xhat-\that)\right]^h, & t > |x|, \end{cases} \\
\Delta T_4(t,x) &= \begin{cases} 2i\sin(2\pi h) e^{2 h(\xhat+\that)} \left[\csch^2(\xhat-\that)\right]^h \pFq{3}{2}{h,h,2h}{1+h,1+h}{e^{2(\xhat+\that)}}, & t < -|x|, \\
		\Theta(x) \frac{2i \Gamma(1+h)^2}{\Gamma(2h)} \left[\pi \cos(\pi h) + 2\sin(\pi h)(\xhat+\that) \right] \left[\csch^2(\xhat-\that)\right]^h, & -|x| < t < |x|, \\
		 -2i \left\{\sin(2\pi h) e^{-2 h(\xhat+\that)} \pFq{3}{2}{h,h,2h}{1+h,1+h}{e^{-2(\xhat+\that)}} \right. & \\ \hfill \left. + \frac{\Gamma(1+h)^2}{\Gamma(2h)}\left[2\sin(\pi h) (\xhat+\that) - \pi \cos(\pi h)\right]\right\} \left[\csch^2(\xhat-\that)\right]^h, & t > |x|. \end{cases}
\eea
\end{subequations}
Combining these discontinuities according to~\eqref{eq:DeltaG2T}, we find that~$\Delta G_2(t,x) = 0$ when~$-|x| < t < |x|$, which justifies the claim made in Section~\ref{subsec:Fourier} that the contour of integration for the Fourier transform in~$t$ can be deformed as shown in Figure~\ref{subfig:tmodifiedcontour}.  On the other hand, when~$t > |x|$ only~$\Delta T_3$ and~$\Delta T_4$ contribute to~$\Delta G_2$, yielding the discontinuity that needs to be integrated over in the Fourier transform:
\begin{multline}
\label{eq:DeltaG2}
\Delta G_2(t,x) = \frac{1}{2} \Theta(t-|x|) \left(\frac{2\pi}{\beta}\right)^{4h} \frac{\pi^2 (1-h)\csch^{2h}(\that-\xhat)}{2^{2h-1} i} \\ \times \left\{\sin(2\pi h) e^{-2h(\xhat+\that)} \left[2(h+1) \coth(\that-\xhat) \pFq{2}{1}{h,2h}{1+h}{e^{-2(\xhat+\that)}} \right. \right. \\ + \left. \left. \frac{(h-1) (1+2h\cosh^2(\xhat-\that))}{h \sinh^2(\xhat-\that)} \pFq{3}{2}{h,h,2h}{1+h,1+h}{e^{-2(\xhat+\that)}}\right] \right. \\ \left. + \frac{\Gamma(h)\Gamma(1+h)}{\Gamma(2h)}\left[2(h+1) \sin(\pi h) \coth(\xhat-\that) \phantom{\frac{1}{2}} \right.\right. \\ \left.\left. + \frac{(h-1)(1+2h \cosh^2(\xhat-\that))}{\sinh^2(\xhat-\that)} \left(2(\xhat+\that)\sin(\pi h) - \pi \cos(\pi h)\right)\right]\right\}.
\end{multline}

\subsection{Contribution from pole}
\label{subapp:Fpole}

The contribution to the Fourier transform in~$t$ from the pole,~$F_\mathrm{pole}$, is given in~\eqref{subeq:Fpole}.  To compute it, we expand~$G_2^\text{E}(\Sigma(t,x),\overline{\Sigma}(t,x))$ around~$t = x$ by writing~$t = x + \eps e^{i\theta}$, expand around~$\eps = 0$ keeping careful track of branch cuts, and then integrate the result in~$\theta$.  Restricting to~$h \in (0,1/2)$ for simplicity, the result is
\begin{multline}
\label{eq:Fpolegeneral}
F_\mathrm{pole}(\omega,x) = -\Theta(x) \frac{e^{i \hat{\omega} \xhat}}{2^{2h}} \left\{\frac{4 \pi^3 (1-h)^2 \Gamma(1+h)}{\Gamma(-h)\Gamma(1+2h) \eps^{1+2h}} \right. \\ \left. \times \left[\frac{2 \Gamma(2h) \cos(\pi h) e^{-4h\xhat}}{\Gamma(1+h)^2} \pFq{3}{2}{h,h,2h}{1+h,1+h}{e^{-4\xhat}} + 4\xhat - \pi \cot(\pi h)\right] \right. \\ \left.
	- \frac{\pi^2(h-1) \sin(2\pi h)}{h \eps^{2h}} \left[2(1+h) e^{-4h\xhat} \pFq{2}{1}{h,2h}{1+h}{e^{-4\xhat}} \right. \right. \\ \left. \left. + \frac{(1-h)(1+2h) e^{-4h\xhat}}{h} \left((2h-i\hat{\omega}) \pFq{3}{2}{h,h,2h}{1+h,1+h}{e^{-4\xhat}} \right. \right. \right. \\ \left. \left. \left. + \frac{4h^3 e^{-4\xhat}}{(1+h)^2} \pFq{3}{2}{1+h,1+h,1+2h}{2+h,2+h}{e^{-4\xhat}}\right) \right. \right. \\ \left. \left. + \frac{\sqrt{\pi} \Gamma(1+h)}{2^{2h} \Gamma(h+1/2) \cos(\pi h)} \left(-4(1+h-h^2) + i \hat{\omega} (1-h)(1+2h)(\pi \cot(\pi h) - 4\xhat) \right)\right] \right. \\ \left.
	- \frac{i\hat{\omega} 2^{1-2h} \pi^3 \Gamma(1/2-h)\Gamma(h)}{\Gamma(1-h)\Gamma(1/2+h)} \left(1+h-h^2 +2i\hat{\omega} h(1-h) \xhat\right) + O(\eps^{1-2h}) \right\},
\end{multline}
where we have defined~$\omegahat = \omega \beta/\pi$.  The omitted terms are~$O(\eps^{1-2h})$ and therefore vanish when~$h \in (0,1/2)$.  The two singular terms, of order~$\eps^{-(1+2h)}$ and~$\eps^{-2h}$, will need to cancel out with the contribution from~$F_\mathrm{cut}(\omega,x)$.

\bibliographystyle{jhep}
\bibliography{bibliography}

\end{document}